\documentclass[fleqn,usenatbib]{mnras}

\usepackage{newtxtext,newtxmath}

\usepackage[T1]{fontenc}

\DeclareRobustCommand{\VAN}[3]{#2}
\let\VANthebibliography\thebibliography
\def\thebibliography{\DeclareRobustCommand{\VAN}[3]{##3}\VANthebibliography}

\usepackage{graphicx}	
\usepackage{amsmath}	

\newcommand{\skipthis}[1]{}

\title[Low-Frequency HII Regions]{A Low Frequency Pilot Survey of Southern H{\sc ii} Regions in the Vela Constellation}
\author[C.D. Tremblay et al.]{
Chenoa D.Tremblay,$^{1}$\thanks{E-mail: astrochenoa@gmail.com}
Tyler L. Bourke,$^{2}$
James A. Green,$^{1}$
John M. Dickey,$^{3}$
O. Ivy Wong,$^{1,4}$ \and
and Tim J. Galvin$^{1,5}$
\\
\\
$^{1}$CSIRO, Space and Astronomy, Australian Telescope National Faciliy, PO Box 1130, Bentley WA 6102, Australia\\
$^{2}$SKA Observatory, Jodrell Bank, Macclesfield SK11 9FT, United Kingdom\\
$^{3}$School of Maths and Physics, University of Tasmania, Hobart, TAS 7001, Australia\\
$^{4}$ International Centre for Radio Astronomy Research, UWA, 35 Stirling Hwy, Crawley, WA 6009, Australia\\
$^{5}$International Centre for Radio Astronomy Research, Curtin University, GPO Box U1987, Perth WA 6845, Australia
}

\date{Accepted XXX. Received YYY; in original form ZZZ}

\pubyear{2021}

\begin{document}
\label{firstpage}
\pagerange{\pageref{firstpage}--\pageref{lastpage}}
\maketitle

\begin{abstract}
Atomic ionised regions with strong continuum emission are often associated with regions of high-mass star formation and low-frequency ($<$2\,GHz) observations of these regions are needed to help build star formation models. The region toward the Vela Supernova Remnant is particularly interesting as it is a complex structure of recent supernova explosions and molecular clouds containing a number of H{\sc ii} regions that are not well characterised.  We searched publicly available catalogues for H{\sc ii} regions, both candidate and identified, which also have low-frequency emission. In the area of $\sim$400 square degrees toward the Vela Supernova remnant, we found 10 such H{\sc ii} regions, some of which have multiple components in catalogues. In this work we use data from the Australian Square Kilometre Array Pathfinder and previously unpublished data from the Murchison Widefield Array and the Australian Telescope Compact Array to analyse these sources. The high-mass star forming region RCW 38, with observations specifically targeted on the source, is used as a pilot study to demonstrate how low-frequency, wide-field continuum observations can identify and study H{\sc ii} regions in our Galaxy. For the 9 other H{\sc ii} regions, we discuss their properties; including information about which clouds are interacting, their ages, whether they are dominated by infrared or optical H$\alpha$ lines, distances, ionising photon flux, and upper limits on the infrared luminosity. In future work, these 9 regions will be analysed in more detail, similar to the result for RCW 38 presented here. 
\end{abstract}




\section{INTRODUCTION }
\label{sec:intro}

H{\sc ii} regions are characterised as areas where interstellar atomic hydrogen is ionised and where star formation has recently taken, or is, taking place. The far-ultraviolet (FUV) radiation from high-mass stars (spectral type late B or O stars, $>$10\,M$_{\odot}$) embedded in their natal environment create a mechanism for ionisation through feedback \citep{Samal_2018}.  This ionisation is strongly correlated with: H$\alpha$ emission (although sometimes only weak emission is detected) readily observable by optical telescopes \citep{Seon-2011}; infrared emission of the warm ionising medium \citep{Anderson_2012}; recombination lines often observed by radio telescopes \citep{Hoglund_Mezger_1965,Wenger_2021}; and sometimes radio continuum emission (i.e. \citealt{Bania_2010,Wenger_2021}).

Within the H{\sc ii} region, we do not know if the low-mass stars or the high-mass stars are the first to form \citep{Kumar_2020}.  This creates some challenges on observing the sources and impacts which observational methods are most likely to detect the nebulae. The densest, brightest regions called ``compact H{\sc ii} regions" (or ``ultra-compact") are often heavily obscured by dust and cannot be viewed by visible light, making infrared and radio measurements crucial.  \cite{Masque_2020} studied ultra-compact H{\sc ii} regions and found that nebulas have a mix of thermal and non-thermal emission. They suggest that the thermal sources have free-free emission from the ionised gas and the non-thermal sources have a nature that appears to be different from standard expanding or in-fall scenarios.  So observations across the electromagnetic spectrum, where either free-free emission or synchrotron radiation dominate, of the same H{\sc ii} clouds are required to help build complete star formation models.

The Vela constellation contains a giant molecular cloud along the Carina–Sagittarius Spiral Arm of the Galactic Plane. The highly energetic environment caused by recent supernova explosions in the near and far-field allow for some unique H{\sc ii} regions. This makes it a particularly interesting target for wide-field surveys.

Although a number of large southern-sky surveys have been completed over the last 40 years, new surveys with current telescope technology has provided significant insight.  \cite{Hindson_2016} completed a low-frequency survey for H{\sc ii} regions with the Murchison Widefield Array (MWA) at $260^{\circ} < \ell < 340^{\circ}$, $-3^{\circ} < b < 2^{\circ}$. This region approaches the IRAS radio shell of the Vela supernova remnant but does not extend beyond or around the remnant. \cite{Wenger_2021} recently completed a blind census of H{\sc ii} regions using the Australian Telescope Compact Array (ATCA) at 4.8\,GHz as part of the ``Southern H{\sc ii} Region Discovery Survey (SHRDS)". In their survey they detected hydrogen radio recombination lines (RRLs) towards 208 previously known nebulae and 438 H{\sc ii} candidates along $259^{\circ} < \ell < 346^{\circ}$, $|b| < 4^{\circ}$. Some of the candidate regions had recombination line detections but no detectable radio continuum emission at 4.8\,GHz.

Often Galactic H{\sc ii} regions are identified through the warm dust emission observed by infrared telescopes or based on their morphology. Radio observations $<$10\,GHz focus on young or compact H{\sc ii} regions (e.g. CORNISH; \citealt{Hoare_2012}).  Many of the surveys of the Galactic Plane have focused closer to the Galactic Centre and many of the low-frequency surveys suffer from low resolution (i.e. \citealt{Green_1999,Hindson_2016}).  With a combination of the new  wide-field and high-resolution capabilities of the Australia Square Kilometre Array Pathfinder (ASKAP) and the Murchison Widefield Array (MWA; with its Phase II upgrade) we can efficiently identify H{\sc ii} regions and provide information on their morphological and physical characteristics, especially when combined with infrared and optical surveys.

In this paper we utilise new and archival data to extend the knowledge of H{\sc ii} regions at $259^{\circ} < \ell < 275^{\circ}$, $-6^{\circ} < b < 4^{\circ}$, some of which is shown in Figure \ref{full}.  The observations from ASKAP and the MWA represent a single wide-field footprint from each telescope.  We have broken the analysis in the paper into two parts. In the first part we provide an in-depth analysis of RCW 38 (Section 4). This source was chosen for extra attention as it is the densest (in stellar membership) embedded region of high-mass star formation within 2 kpc after the Orion Nebula Cluster (ONC), but is not as well studied \citep{Wolk_2008,Lada_2003}.  In the second part we compare catalogues of known and candidate H{\sc ii} regions from the WISE Catalogue of Galactic H{\sc ii} Regions V2.3 \citep{WISE_Catalogue}\footnote{http://astro.phys.wvu.edu/wise/} and from SHRDS. These were overlaid on the ASKAP and MWA images to  search for sources that show low-frequency (114\,MHz and/or 887.5\,MHz) emission (Section 5). For each region that shows radio continuum emission at these frequencies, we provide a short analysis. 

\section{Observations}
\begin{figure*}
	\includegraphics[width=0.98\textwidth]{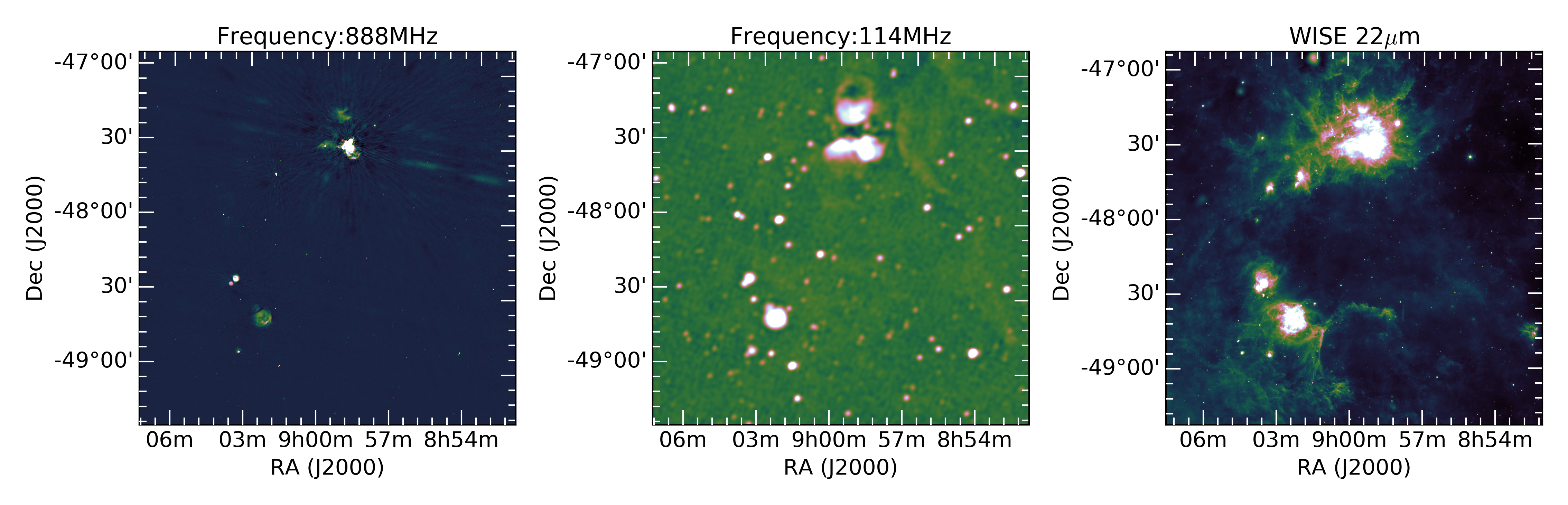}
	\caption{An image showing a region covered by new data from Rapid ASKAP Continuum Survey (left) at 887.5\,MHz and MWA image at 114\,MHz (Middle). The striations around RCW 38 and other bright sources in the ASKAP image are due to the shallow, 15 minute integrations and small calibration errors ($<$1\,per\,cent;\citealt{McConnell_2020}) which are obvious around RCW 38 as is the source is bright.  The WISE 22$\mu$m (right) shows the infrared emission and highlights some of the H{\sc ii} regions discussed in this work as marked on the MWA image.}
\label{full}
\end{figure*}

New observations centred on the Vela Supernova Remnant were completed using the MWA and matched with archival observations from ASKAP and WISE (at 22\,$\mu$m) as shown in Figure~\ref{full}.  Both the ASKAP and MWA telescopes are located in the Murchison Radio-astronomy Observatory (MRO), the site designated for the low-frequency component of the future Square Kilometre Array (SKA).  We also used previously unpublished data at 1.6 GHz of RCW 38 from the Australian Telescope Compact Array (ATCA).  

\subsection {Murchison Widefield Array (MWA)}
Using the Murchison Widefield Array \citep{Tingay_2013,Wayth_2018, Beardsley_2019} we observed the region of the Vela and Puppis Constellation for 30\, hours between 2018 January 05 -- 23, of which 17 hours are used in this analysis.  The remaining data suffered significant artefacts (visible in the $u,v$ data as deformations of the beam, likely caused by a malfunction in the electronics during the extended array deployment) or radio frequency interference and were not used for analysis. These observations represent early science data taken during the Phase II upgrade \citep{Wayth_2018,Beardsley_2019} and use 91 dipole tiles with baseline lengths ranging from 22\,m to 5.1\,km at a central frequency of 114\,MHz.

The MWA correlator offers 24 $\times$ 1.28\,MHz coarse channels across a bandwidth of 30.72\,MHz.  The data are calibrated and imaged as described in \cite{Tremblay_2020_MWA}.  For each night, 2-minute observations of the galaxy Hydra A (245.8$\pm$31.9 Jy at 160\,MHz; \citealt{1981A&AS...45..367K}) are used to perform primary bandpass and phase calibration.  The solutions are further refined by using self calibration prior to being applied to each set of visibilities.  A robust weighting with a factor of ``0.5'' was used for the images presented here, resulting in a synthesised beam of $129\arcsec\ \times\ 105\arcsec$ at $-$29$\degr$ position angle.

The Earth's atmosphere can create spatially variant refraction and propagation delays that are more evident at lower frequencies.  For each 5-minute observation, the source positions are compared to the Molonglo Reference Catalogue (MRC; \citealt{Large_1981}) to correct for shifts in the source positions.  After correction, the error in a stacked continuum image was --10$\pm$26\,arc\,seconds in Right Ascension and 6$\pm$11\,arc\,seconds in Declination.  Both of these values are smaller than the synthesised beam of 1.03\,arc\,minute.  The ratio of integrated intensity and peak intensity, determined from point sources, is 1.05, suggesting that the ionospheric distortions are well corrected for in these observations.  

\subsection{Australian Telescope Compact Array (ATCA)}
The Australian Telescope Compact Array (ATCA) is a 6-element predominantly East-West interferometer\footnote{it has a small northern spur available for some compact configurations} with 22\,m antennas operating at cm and mm wavelengths.  To complement Parkes single-dish OH observations published in \cite{Bourke_2001}, RCW 38 was observed at 1666\,MHz with the ATCA in its 1.5D and 6.0C configurations in 1996 May and June (preliminary results are presented in \cite{Bourke_2004}).  On 1996 May 10 a 13.7 hour track was obtained with the 1.5D configuration array, while on 1996 June 15 a 11 hour track was obtained with the 6.0C configuration. In order to obtain the highest spectral resolution available with the correlator available at the time of the observations, while still being able to observe both the 1665 and 1667 OH transitions, a 4\,MHz bandpass with 4096 channels centred on 1666.0\,MHz with one linear polarisation was used, resulting in a resolution of 976.56\,Hz or 0.176\,km\,s$^{-1}$. With the correlator available at ATCA at the time these observations were taken, correlating all polarisation products – XX*, YY*, XY*, YX* (where * denotes the complex conjugate) – from the two orthogonal linear polarisations, X and Y, would result in a decrease in the spectral resolution by a factor of 4, resulting in a minimum spectral resolution of 0.8\,km\,s$^{-1}$ at the OH 1.6\,GHz frequencies.  However, the full-width-half-maximum (FWHM) of the Gaussian component responsible for the observed OH Zeeman effect in RCW 38 using the Parkes Radio telescope is only 2.4\,km\,s$^{-1}$ \citep{Bourke_2001},.  
Thus, any OH Zeeman experiment on RCW 38 with ATCA at the time of the observations using both OH lines would result in only 3 spectral channels across the FWHM of this Gaussian component, which is insufficient to derive and fit the Stokes V spectrum.  Therefore, the observations were restricted to only one polarisation with the full spectral resolution, in order to study the kinematics of the OH .

The primary flux calibrator used was PKS 1934--638 with an assumed flux of 14.16\,Jy and the complex gain calibrator was 0823--500.  The visibility data was processed in through standard methodologies using the MIRIAD \citep{Miriad} package.  Bandpass solutions were determined for PKS 1934--638 and copied to PKS 0823--500.  The time dependent antenna based gains were determined for 0823--500, bad data were flagged, and the flux scale was bootstrapped to 1934--638.  The gain solutions were copied to RCW 38, the continuum was constructed from the line-free channels in the visibility data set and the line and continuum visibilities split.  Very little data needed to be flagged and very little interference was observed ($<<$10\,\%).  The data for the two configurations were calibrated separately and combined at the imaging stage.  Maps were made by Fourier transforming the visibilities, using the {\sc CLEAN} algorithm down to the rms (root mean squared) noise level, and restoring the image using a Gaussian beam.  A robust weighting with a factor of ``0.5" was used for the images presented here, resulting in a synthesised beam of 9.8 $\times$ 8.3 arc\,second at 16.8$^{\circ}$ position angle for the line data.  The primary beam size at these frequencies is $\approx$25\,arc\,minutes.

\subsection{Ancillary Data}
In addition to the observations above, we used the public catalogues of 887.5\,MHz Stokes I images from the recent Australian Square Kilometre Array survey named Rapid ASKAP Continuum Survey (RACS;\citealt{McConnell_2020}) in this work. These images were downloaded from The CSIRO ASKAP Science Data Archive (CASDA\footnote{https://research.csiro.au/casda/}) and represent publicly available data sets. We received additional ATCA radio astronomy data from \cite{Wolk_2008} at 4800\,MHz for RCW 38.

Additionally, we used data from the Wide-field Infrared Survey Explorer (WISE), which is a joint project of the University of California, Los Angeles, and the Jet Propulsion Laboratory/California Institute of Technology.  To understand the dust emission we used the 90\,$\mu$m survey from AKARI Far-Infrared All-Sky Survey by \cite{Doi_2015} and Herschel Space Observatory Archive\footnote{http://archives.esac.esa.int/hsa/whsa/}. For the H-alpha emission we used The Southern H-Alpha Sky Survey Atlas (SHASSA; \citealt{Gaustad_SHASSA}), Wisconsin H-Alpha Mapper (WHAM; \citealt{WHAM}), and AAO/UKST SuperCOSMOS H{\ensuremath{\alpha}} survey (SHS;\citealt{Parker_2005_SHS}).  Images from these surveys were obtained through $SkyView$\footnote{https://skyview.gsfc.nasa.gov/} or the individual survey web servers.

We note that the sources were not covered in the MWA Galactic and Extra-galactic all-sky survey catalogue (GLEAM; \citealt{Hurley-Walker_2017,HW_2019_GLEAMGAL}). We do note that one source, RCW 39 was covered by \cite{Hindson_2016} in their survey of H{\sc ii} regions using GLEAM Galactic Plane data.

\section{H{\sc ii} Region Identification}
\begin{figure*}
\centering
\includegraphics[width=0.31\textwidth]{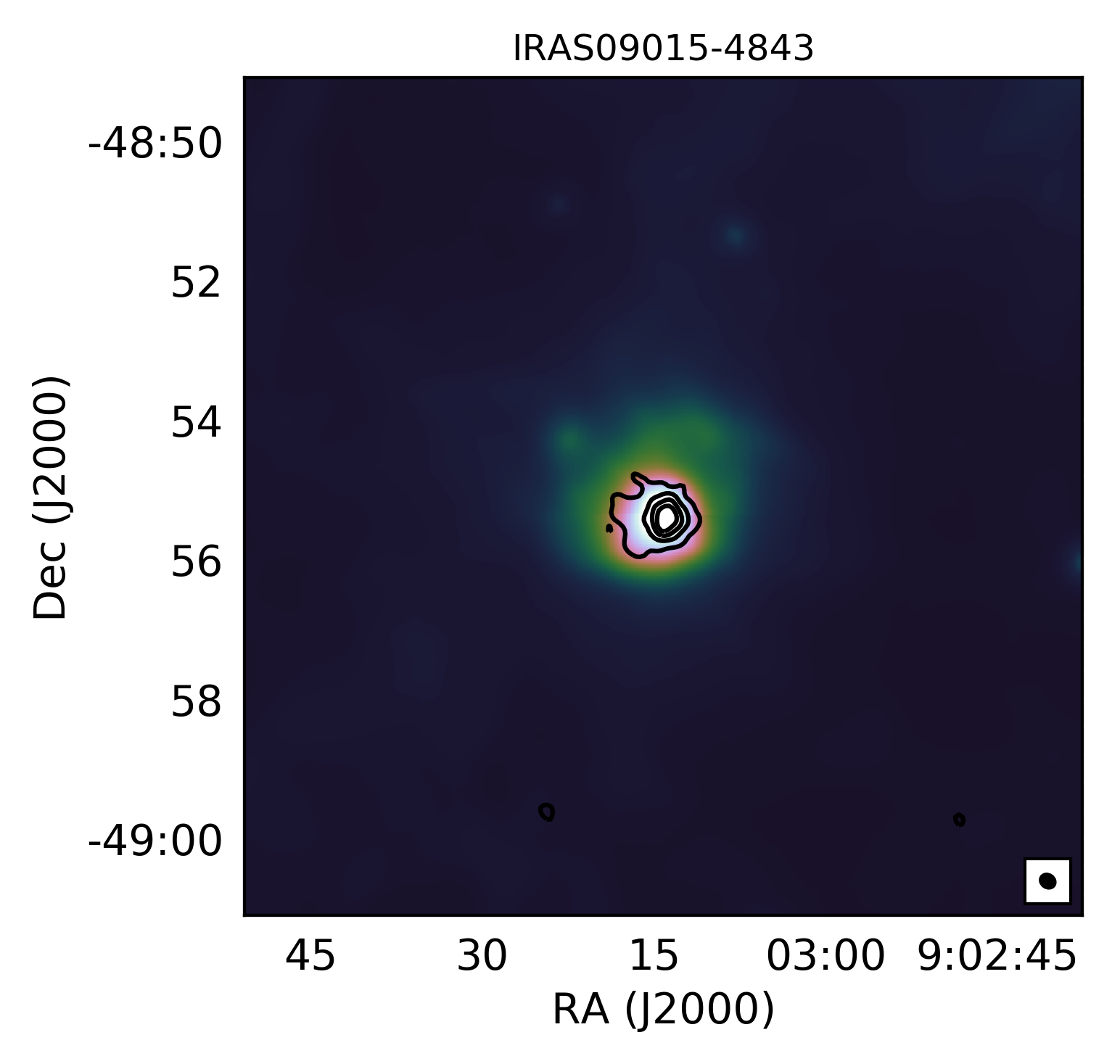}
\includegraphics[width=0.33\textwidth]{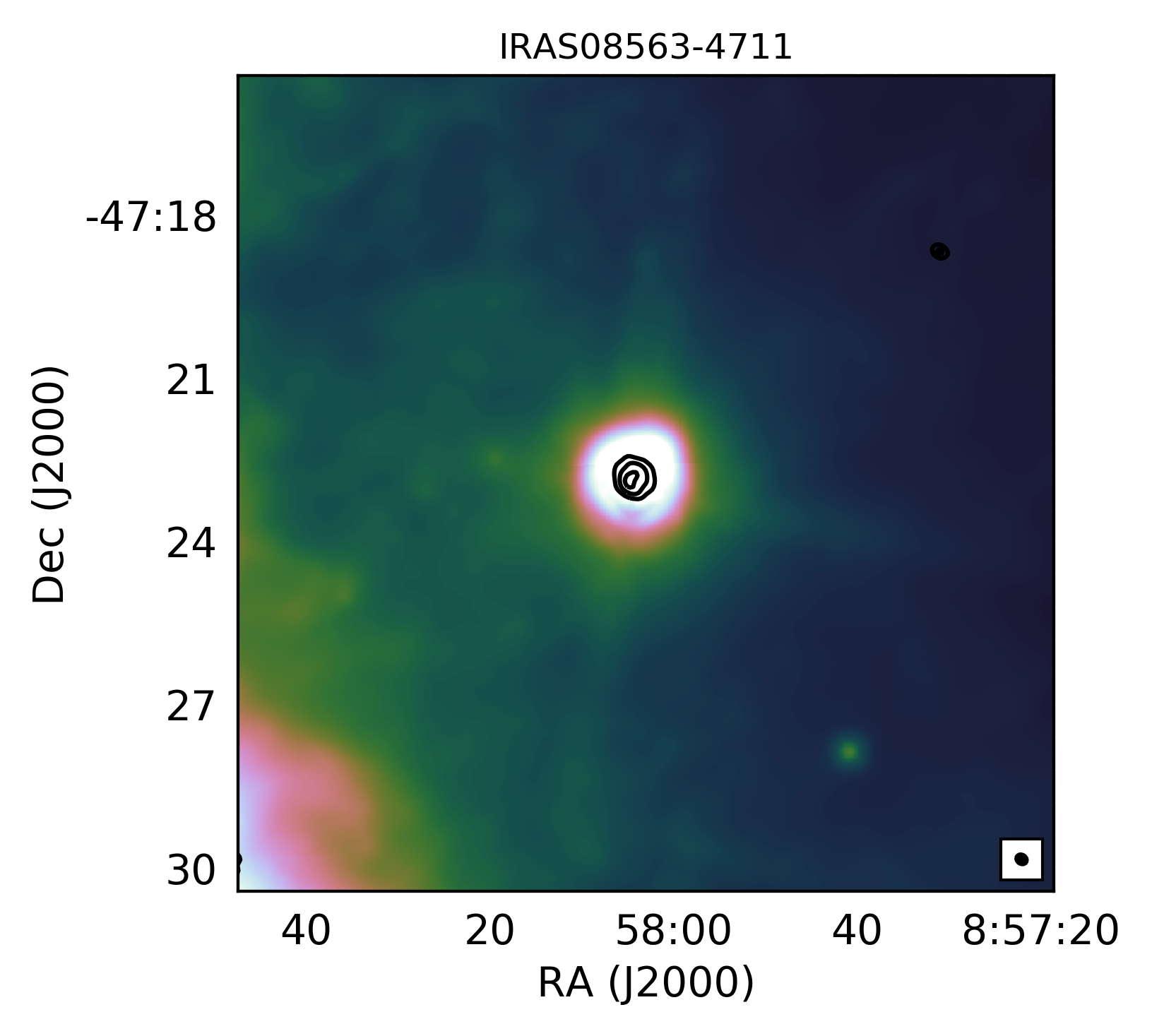}
\includegraphics[width=0.33\textwidth]{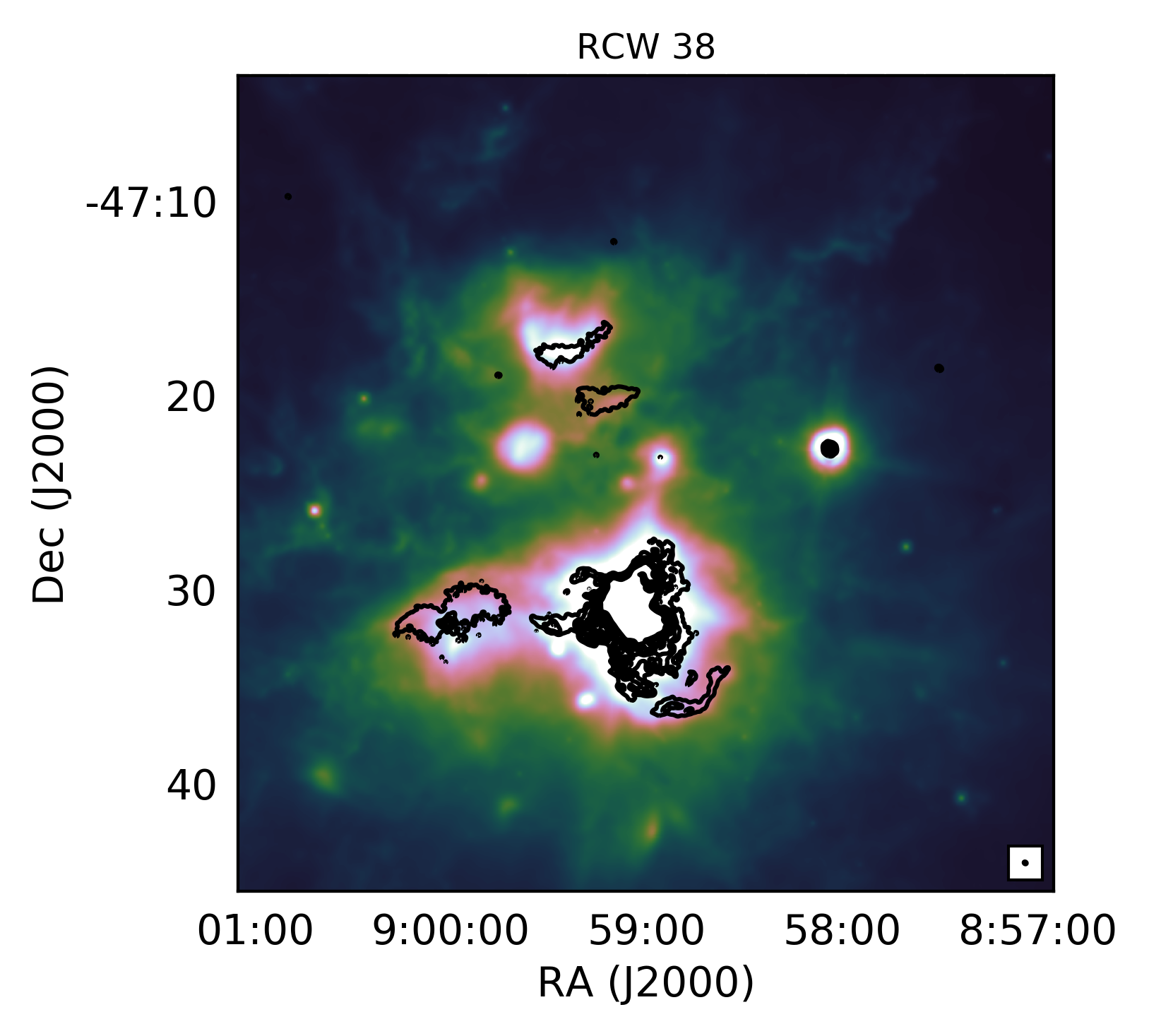}
\includegraphics[width=0.32\textwidth]{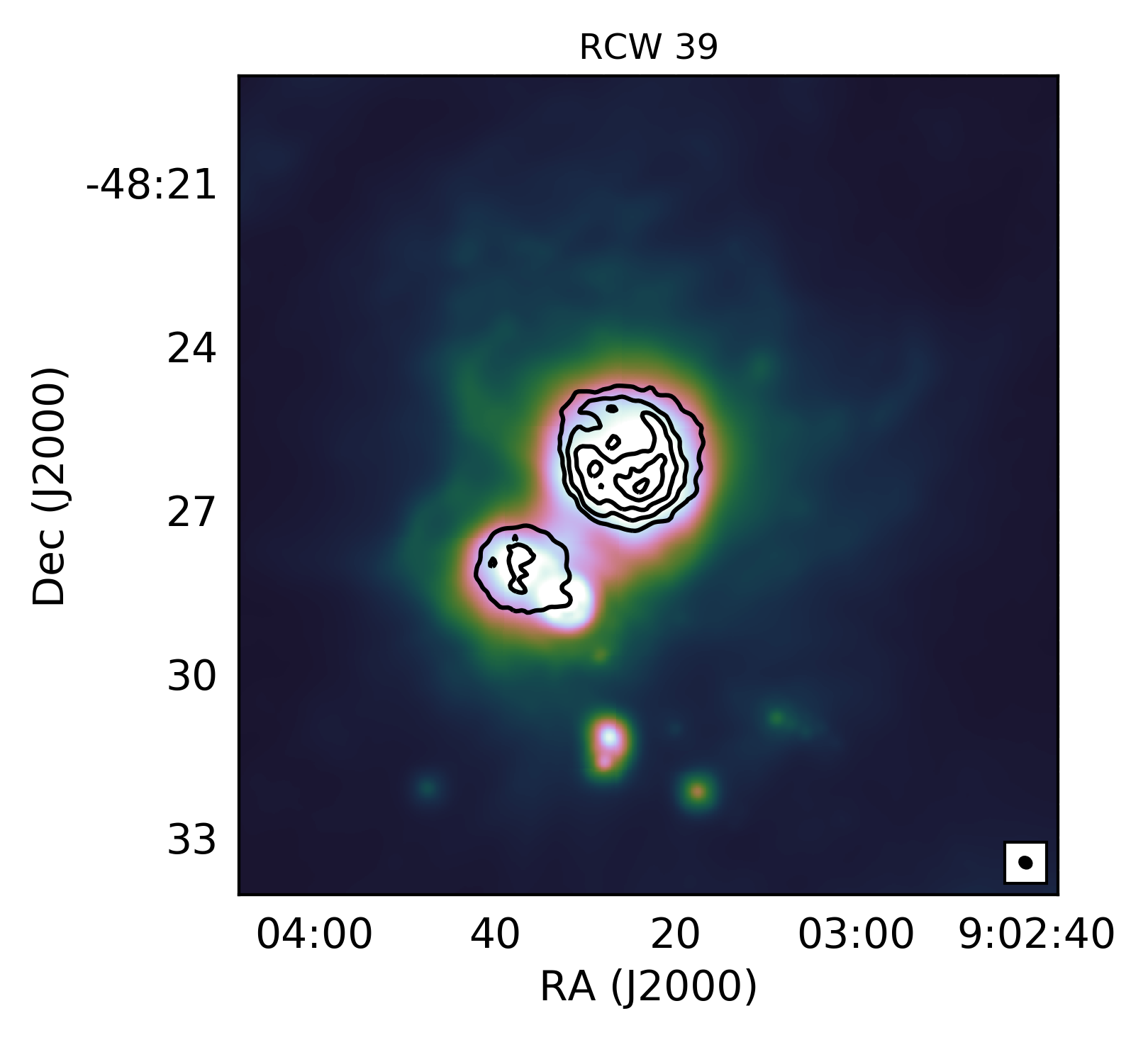}
\includegraphics[width=0.32\textwidth]{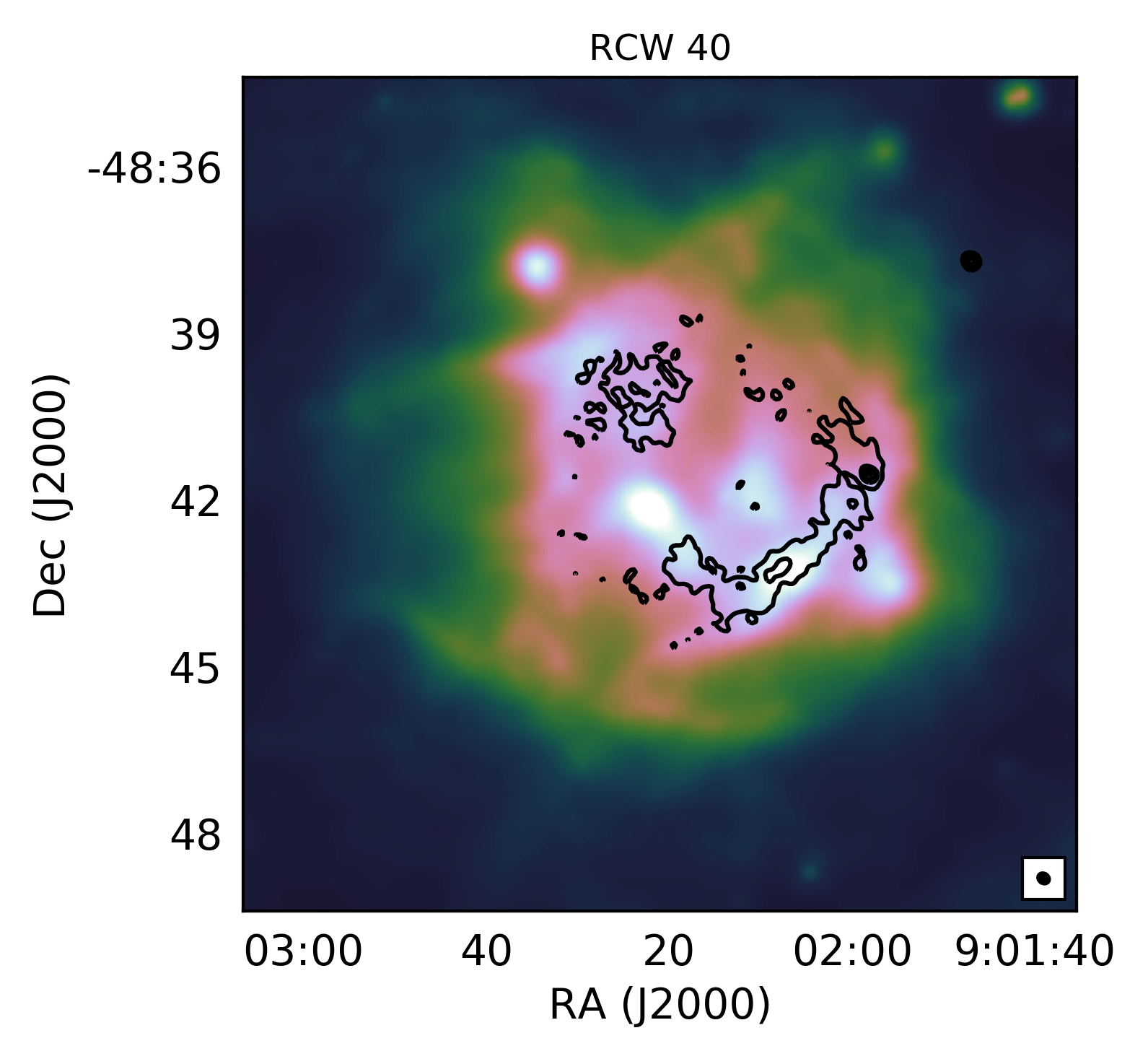}
\includegraphics[width=0.32\textwidth]{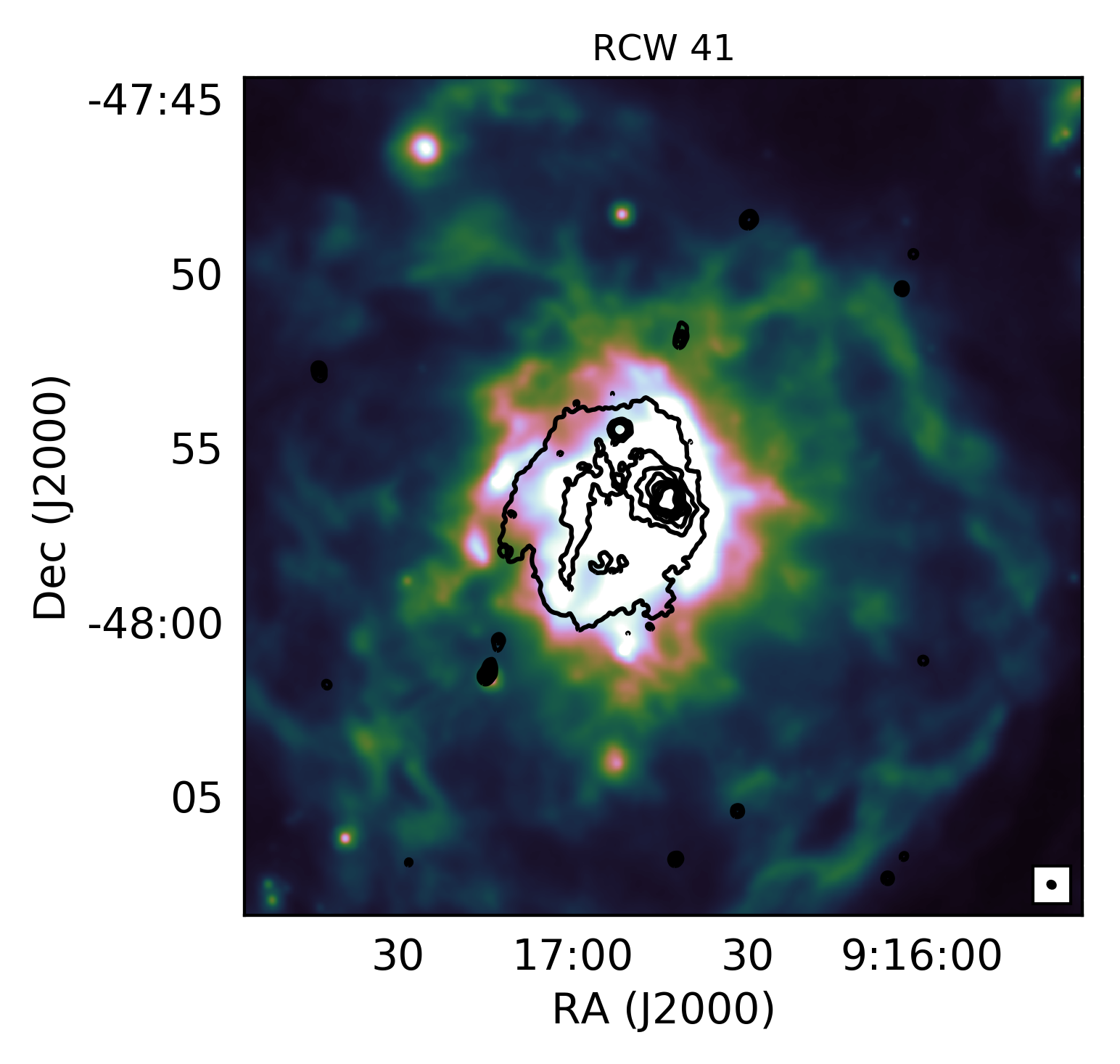}
\includegraphics[width=0.32\textwidth]{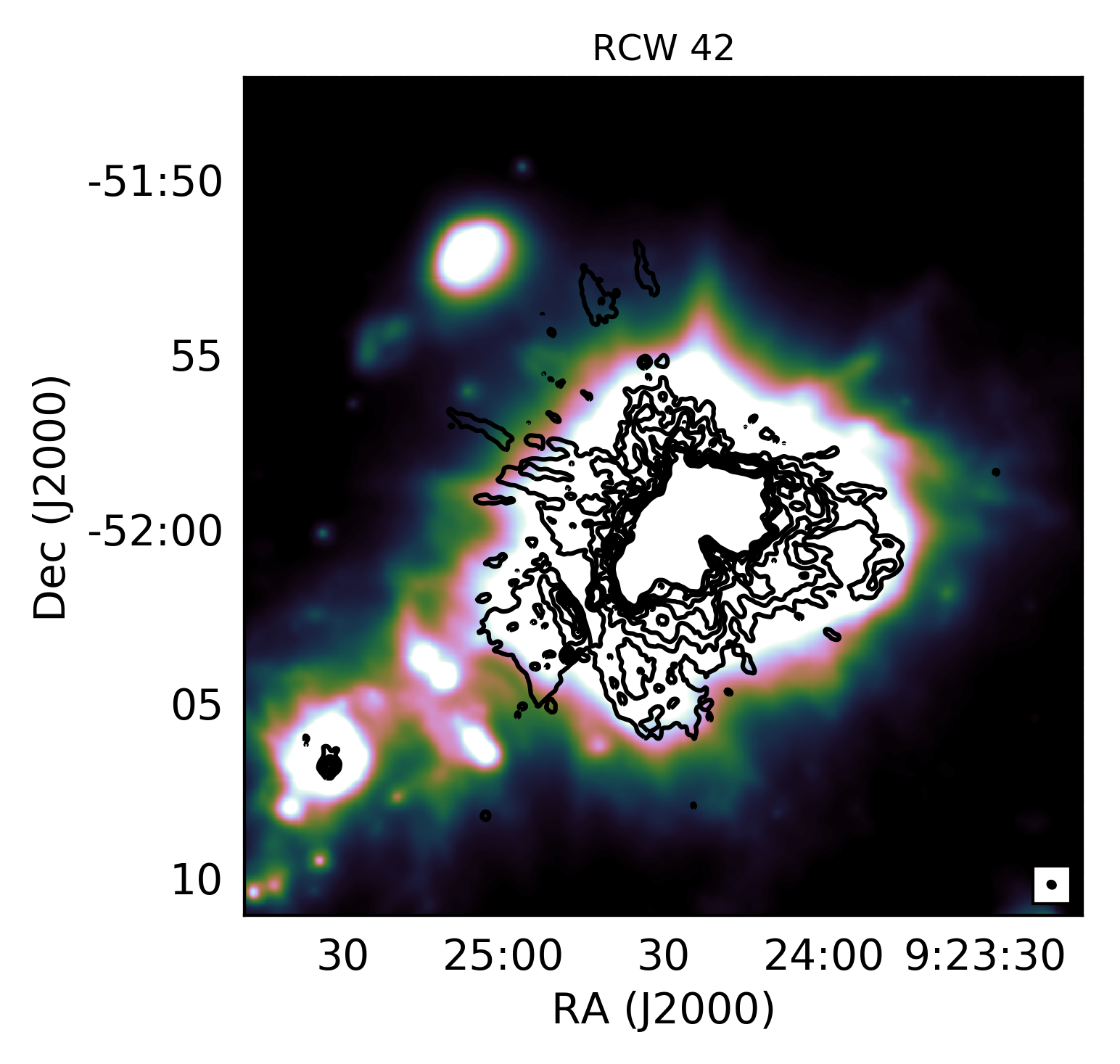}
\includegraphics[width=0.32\textwidth]{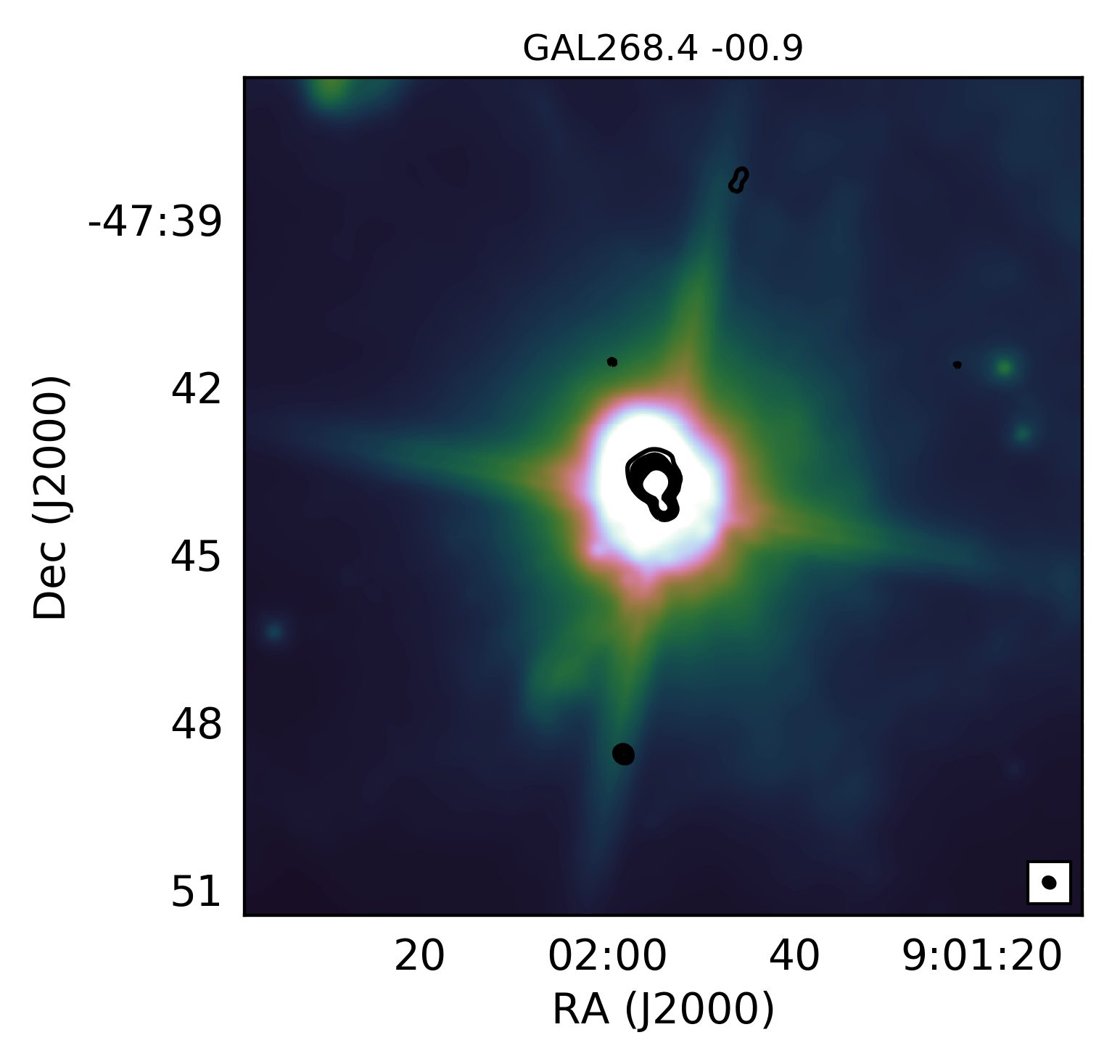}
\caption{The black contours from the ASKAP radio continuum emission are overlaid onto an 22\,$\mu$m infrared image of the region from the Wide-field Infrared Survey Explorer (WISE) survey. The contours show the positions of highest intensity are well matched between the radio and infrared emission. The contours are set at levels of 0.02--0.20\,Jy in increments of 0.03\,Jy and the bottom right-hand corner shows the size of the ASKAP synthesised beam}
\label{WISE_ASKAP}
\end{figure*}

\begin{figure*}
\centering
\includegraphics[width=0.34\textwidth]{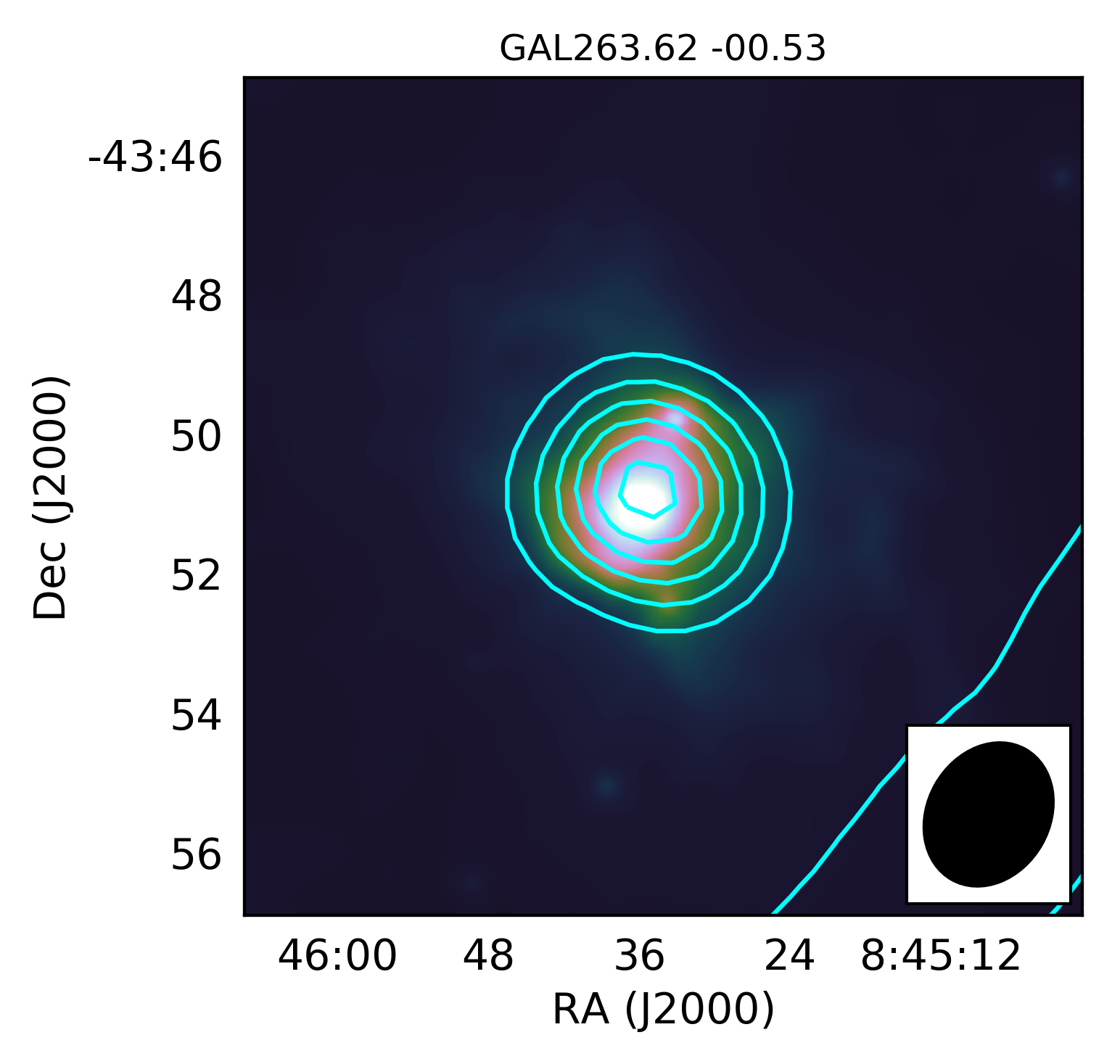}
\includegraphics[width=0.34\textwidth]{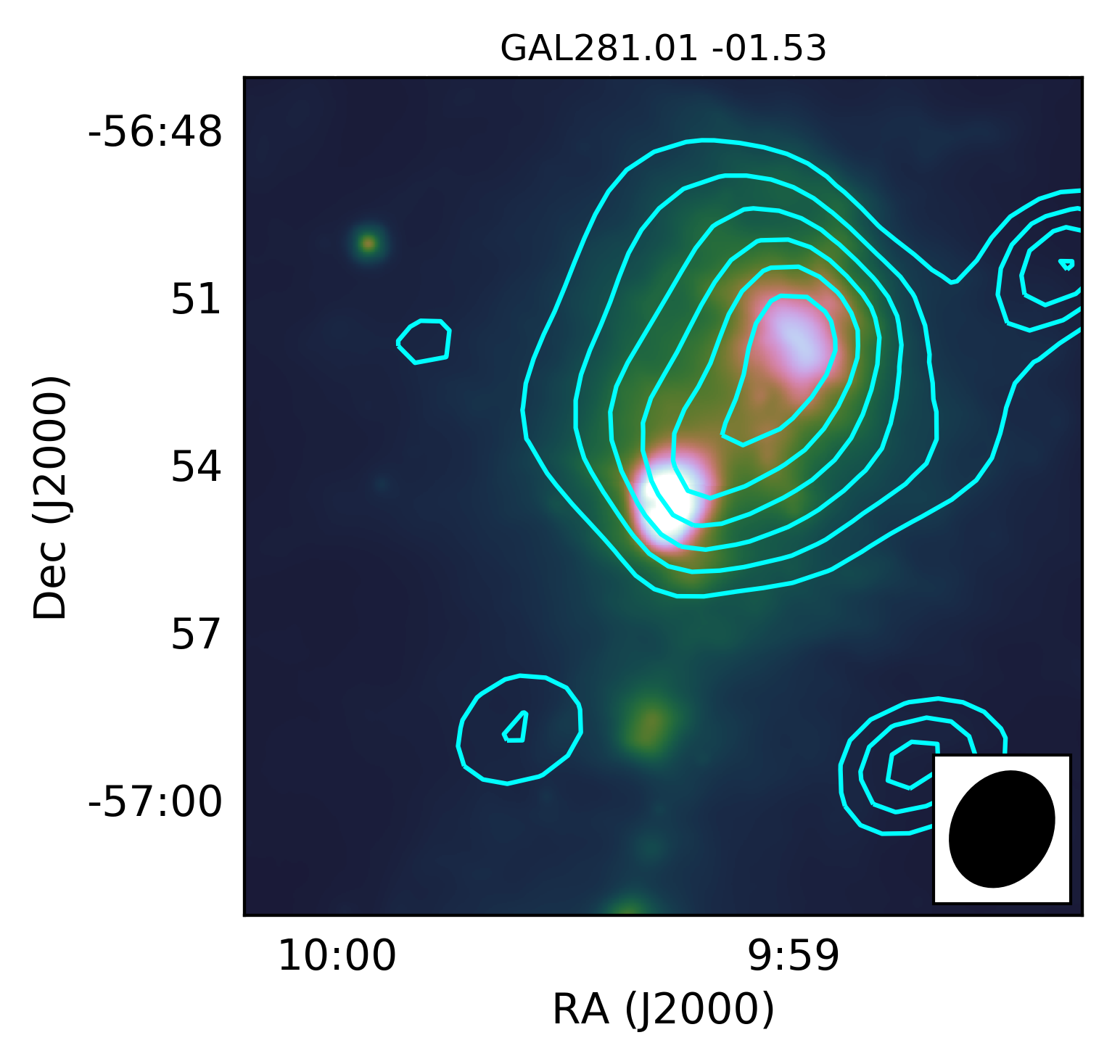}
\caption{The cyan contours from the MWA radio continuum emission are overlaid onto an 22\,$\mu$m infrared image of the region from the Wide-field Infrared Survey Explorer (WISE) survey. The contours show the positions of highest intensity are well matched between the radio and infrared emission. The contours are set at levels of 0.05, 0.1, 0.15, 0.2, 0.25, 0.3\,Jy and the bottom right-hand corner shows the size of the MWA synthesised beam}
\label{WISE_MWA}
\end{figure*}

Atomic ionised regions with strong low-frequency radio continuum emission can be a population of a mix of high-mass star forming regions and planetary nebula (or other dying stars) \citep{ISM_Spitzer, Hoare_2012, Anderson_2012}.  Although, at times these two are grouped into the category of ``H{\sc ii} regions'', as both are composed of gas ionised by photons with energies above the Hydrogen ionisation energy of 13.6\,eV, they represent different phases of stellar life. The standard H{\sc ii} region is ionised by hot spectral type O or early B stars (or clusters of such stars) and associated with regions of recent high-mass ($>$10\,M$_{\odot}$) star formation (e.g. \citealt{Zinnecker_2007}).  Planetary nebula are the ejected outer envelopes of asymptotic giant branch (AGB) stars photoionised by the hot remnant stellar core (e.g. \citealt{Pena_2021}).  There are different methods of distinguishing these two stages, such as the detection of radio recombination lines.  However, as \cite{Hajduk_2021} show, low frequency emission and recombination lines can also be detected in planetary nebula. One way of differentiating sources within these two evolutionary stages is through distinct 22\,$\mu$m infrared emission, which traces the stochastically heated small dust grains \citep{Watson_2008}.  \cite{Calzetti_2007} suggest that the mid-infrared emission traces the dust–obscured star–formation while H$\alpha$ optical emission traces the unobscured one.  As some of the regions are candidates based on recombination line detections, to verify if the sources studied here are high-mass star forming regions, we compare with WISE 22\,$\mu$m emission and the ASKAP continuum emission or MWA continuum emission, as shown in Figures \ref{WISE_ASKAP} and \ref{WISE_MWA}.  

Observations of H{\sc ii} regions below $\sim$200\,MHz can appear optically thick and display a positive spectral index \citep{Hindson_2016}.  The MWA observations used in this paper were completed using the Phase II ``long-baseline array'' with baselines from 22\,m to 5.5\,km.  This means there is low surface brightness sensitivity and a synthesised beam of $\sim$1\,arc\,minute, allowing for the regions to be less confused than the survey by \cite{Hindson_2016}, but with a reduction in the reported flux in comparison to observations with a single-dish telescope.  The observations of the field using ASKAP have a $\sim$13\,arc\,second beam and, at a frequency of 887.5\,MHz,  can be in the optically thin regime. Therefore, the continuum emission from RACS in the regions should be more comparable to WISE, as the resolution of $\sim$12\,arc\,second and a 6\,mJy sensitivity limit are well matched.  

For this survey, the region was searched in $SIMBAD$\footnote{http://simbad.u-strasbg.fr/simbad/}, SHRDS catalogue, and the WISE H{\sc ii} catalogue for known or candidate H{\sc ii} regions.  This cross match found 10 regions which are shown in Figure \ref{WISE_ASKAP} and Figure \ref{WISE_MWA} and summarised in Table \ref{source_list}. All sources have well matched radio continuum and 22\,$\mu$m infrared emission. This, in combination with the detection of radio recombination lines and some masers, confirms the identification of the sources being H{\sc ii} regions containing high-mass stars. Within the literature, a number of these have been characterised as ultra-compact H{\sc ii} regions through their methanol maser emission (e.g. \citealt{Walsh_1997}). In Section 4 and Section 5 we provide analysis of the age and condition of these regions.

\begin{table*}
\small
\caption{Information on the sources observed in this survey and have shown continuum emission at either 114\,MHz or 887.5\,MHz or both. }
\label{source_list}

\begin{tabular}{lcccccc}
\hline

Cluster Name	&	Coordinates	&	SIMBAD Class.	&		Obs. Telescope	&	Maser Detection* & RRL Detection** & Molecular Obs.***	\\
& (J2000)&&&&&\\
\hline
RCW38 & 08 59 05.50 --47 30 39.4& H{\sc ii} Region & MWA \& ASKAP & No Obs. & Yes & CH, OH, $^{18}$CO,\\
&&&&&& $^{12}$CO, CS\\
IRAS08563-4711 & 08 58 04.2 --47 22 57 & H{\sc ii} Region & MWA \& ASKAP & No Detection & Yes& \\
RCW39 & 09 03 21.4 --48 25 48& H{\sc ii} Region & MWA \& ASKAP & No Detection & Yes & \\
RCW40 & 09 02 21.3 --48 41 55 & H{\sc ii} Region & MWA \& ASKAP & No Detection & Yes & \\
RCW42 & 09 24 30.1 --51 59 07 & H{\sc ii} Region & MWA \& ASKAP & No Obs. & Yes & \\
GAL268.4--00.9 & 09 01 54.3 --47 43 59& H{\sc ii} Region & MWA \& ASKAP & No Detection & Yes & \\
IRAS09015-4843 & 09 03 13.5 --48 55 21& H{\sc ii} Region & MWA \& ASKAP & H$_{2}$O,CH$_3$OH(II) & Yes & $^{12}$CO \\
GAL263.62--00.53 & 08 45 34.9 --43 51 07& H{\sc ii} Region & MWA & No Detection & Yes & \\
RCW41 & 09 16 43.2 --47 56 26 & Composite Object & MWA \& ASKAP & H$_{2}$O,CH$_3$OH & No & $^{12}$CO\\
GAL281.01--01.53 & 09 59 08.5 --56 52 44 & H{\sc ii} Region & MWA & No Detection & Yes & $^{13}$CO\\

\hline
\hline
\multicolumn{7}{l}{%
  \begin{minipage}{8.5cm}%
    \small * Information from http://maserdb.net/.\\
** From Southern H{\sc ii} Region Discovery Survey \citep{Wenger_2021} and RCW 38 Study by \cite{Torrii_2021}.\\
***As listed in SIMBAD or WISE Catalogue.\\%
  \end{minipage}%
}\\
\end{tabular}

\end{table*}

\section{RCW 38}
\subsection{Background}
\begin{figure}
    \includegraphics[width=0.45\textwidth]{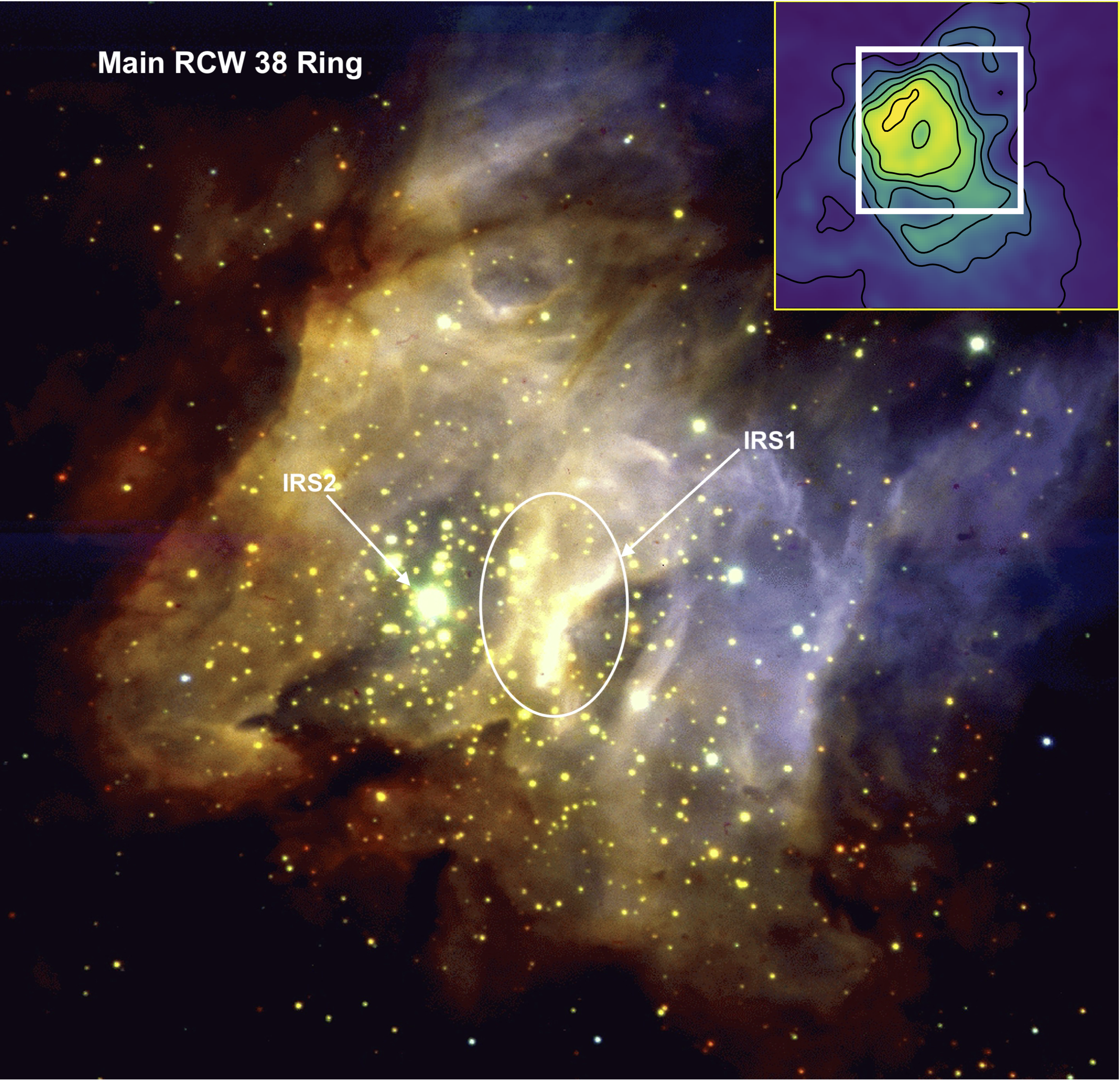}
    \caption{Near-infrared overview of the central 2$\farcs$5 ($\sim$1.2 pc) of RCW 38 main ring centred on RA 08:59:05, Dec --47:30:39. The filters used are Z (blue), H (green) and K (red). The Figure is inspired by Fig. 3 of Wolk et al. (2008) and Fig 1a. of Fukui et al. (2016). The top right-hand corner shows the ASKAP radio continuum image with a white box showing the region which the VLT image covers. The VLT image shows the separation of IRS1 (shock/ionisation front) and IRS2 (O5.5 binary) that is not easily discerned at the radio wavelengths presented here, due to the lack of emission from IRS2. Image Credit: ESO }
    \label{VLT}
\end{figure}

The compact H{\sc ii} region RCW 38 is a relatively nearby region of massive star formation ($\sim$1.63 kpc; \citealt{Zucker_2020}).  It stands out as one of the brightest regions in the southern sky at radio wavelengths \citep{Wilson_1970}, and is one of the few regions within 2\,kpc, along with the Orion Nebula Cluster (ONC), to contain over 2000 young stars identified at infrared wavelengths and in X-rays, although it is more embedded and younger ($\sim$1\,Myr) than the ONC \citep{Wolk2006,Wolk_2008,Winston_2011}.  These studies have identified around 30 OB star candidates, some of which have been confirmed \citep{DeRose2009,Povich2017}. At infrared wavelengths the region is dominated by two sources, the mid-infrared bright RCW 38 IRS1 and the near-infrared bright RCW 38 IRS2 (Figure \ref{VLT}; \citealt{Furniss_1975,Frogel_1974,Smith_1999}).  IRS1 is the peak of an extended emission region on the western side of a ring-like structure of size 0.1\,pc, and is an active site of on-going star formation  \citep{DeRose2009,Winston_2011,Torrii_2021}. Within this ring, near its centre but slightly offset, is IRS2, a binary O5.5 stellar system that is the dominant source of energy for the region, whose winds are responsible for carving out the ring (likely a partial bubble seen in projection) within the natal molecular cloud and clearing out the dust within it \citep{Smith_1999,DeRose2009}, forming the compact H{\sc ii} region.  The ring-like structure, first revealed at mid-infrared wavelengths, has subsequently been observed at wavelengths from the near-infrared through the far-infrared, millimetre, and the radio (e.g., \citealt{Bourke_2004,vigil_2004,Wolk2006,Wolk_2008,Winston_2011,Fukui2016,Torrii_2021,Izumi_2021}). 

The infrared ring is surrounded by dense gas (~10$^{3-4}$ cm$^{-3}$, up to 10$^6$ cm$^{-3}$) on large scales, as observed in e.g. $^{13}$CO 1--0 and CS 2--1, with significant clumps to the north north-east and south south-west \citep{Bourke_2004,Kaneda_2013,Fukui2016}. More compact dense gas traced by CS 2--1 and C$^{18}$O 2--1 is evident to the west and bordering IRS1, indicating that the emission from IRS1 is due to irradiation and ionization of the outer edge of this dense gas \citep{Wang_2016,Torrii_2021}. Observations at mm wavelengths trace both free-free and dust emission; in particular the continuum emission at 1.3\,mm at IRS1 is a mix of both types of emission with a significant free-free component (tens of percent), as indicated by extrapolation of radio fluxes, and the presence of compact mm recombination lines (M. Vigil private communication; \citealt{Wang_2016,Torrii_2021}).

\subsection{Radio observations}
In H{\sc ii} regions, the structure can be traced by emission of optical H$\alpha$ (Balmer lines), infrared emission, and radio continuum. The H$\alpha$ emission traces the heated ionised gas in the diffuse layers of gas and dust while in the compact, dense regions the dust absorbs the radiation and re-emits it in the infrared \citep{ISM_Spitzer}. The continuum emission traces the free-free emission and synchrotron radiation associated with the electron density of the region \citep{Radio_Basics}.

\begin{figure*}
    \includegraphics[width=0.485\textwidth]{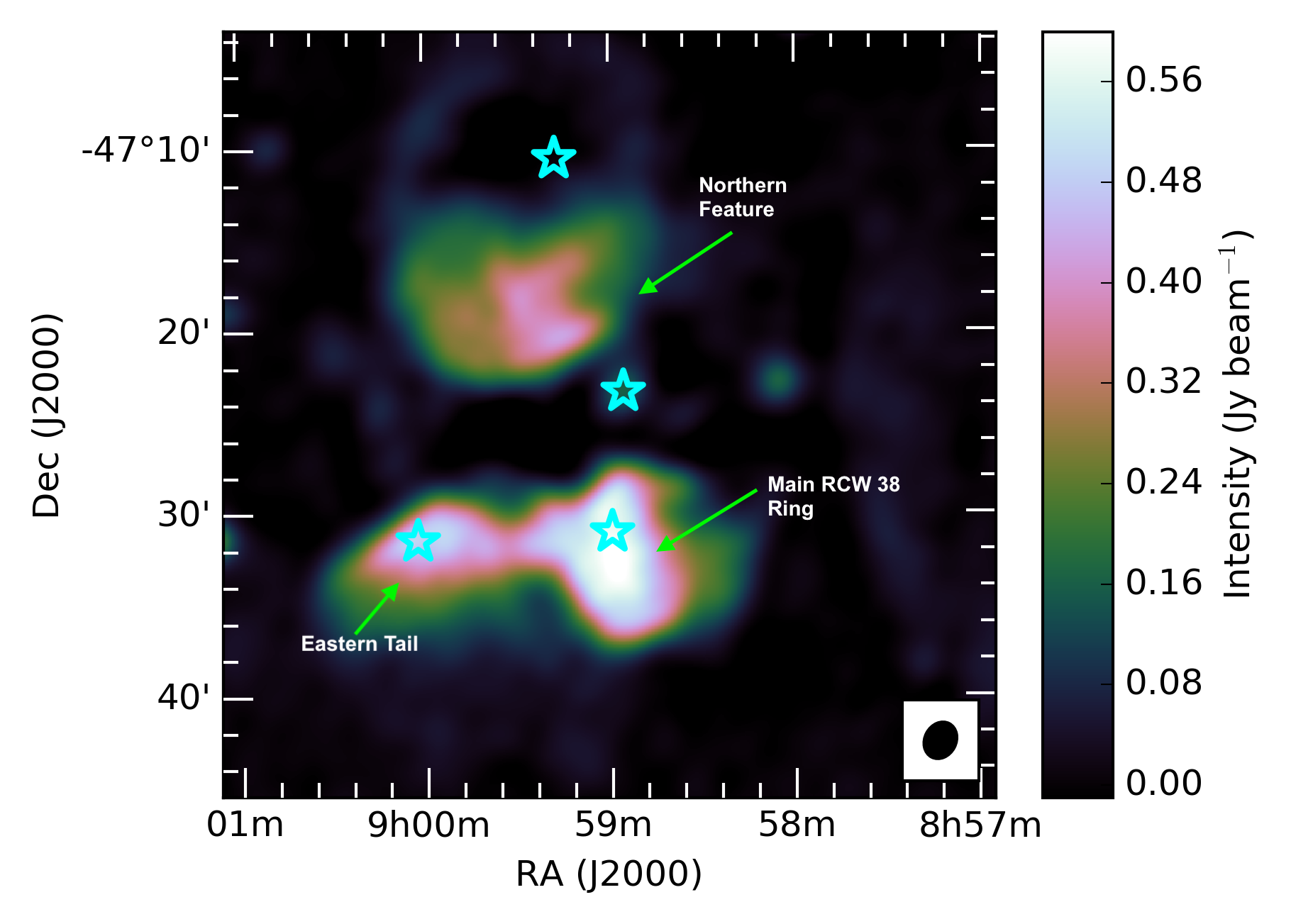}
    \includegraphics[width=0.47\textwidth]{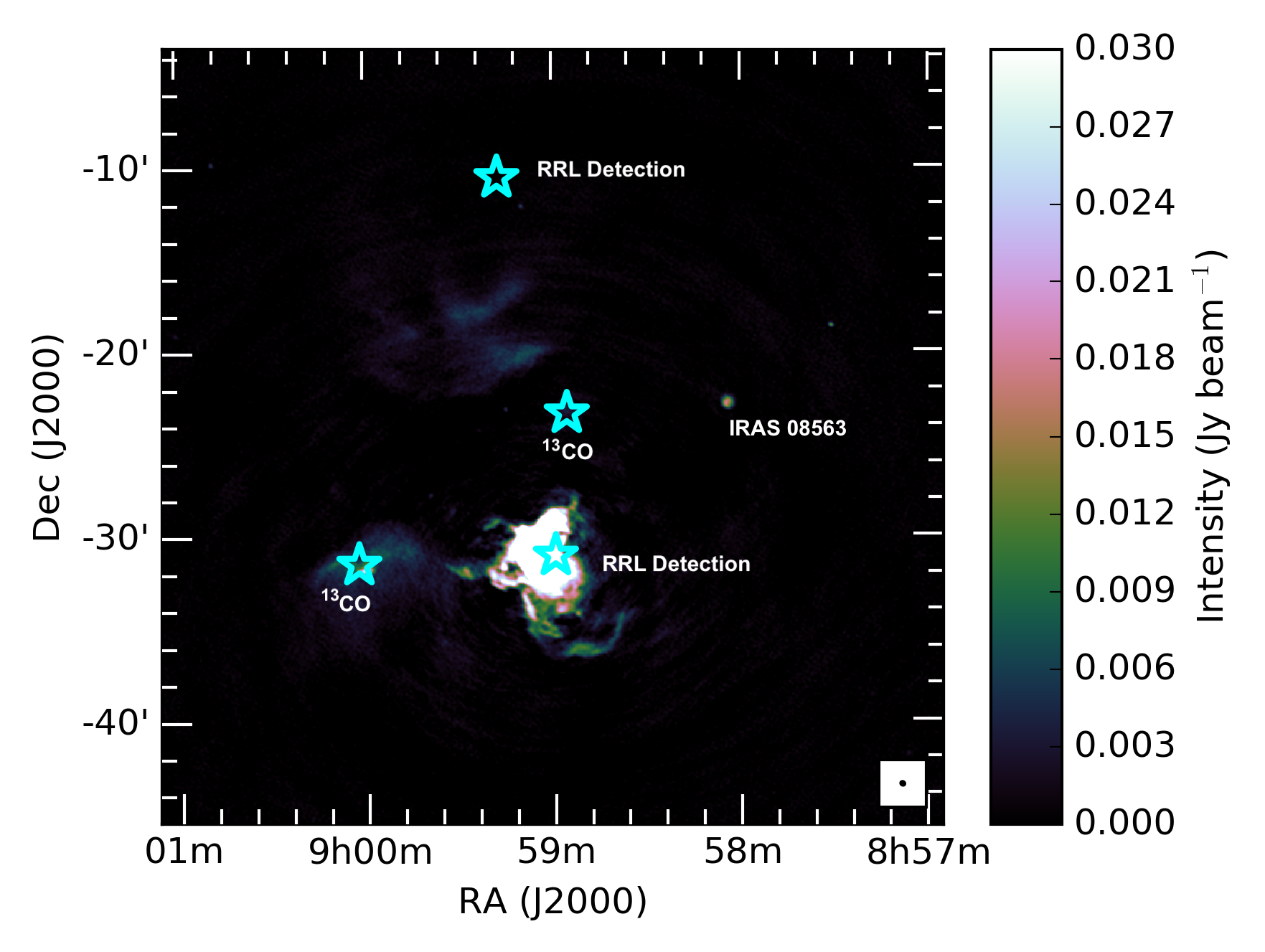}
    \caption{Image of the continuum emission around RCW 38 observed by the MWA at 114\,MHz (left) and ASKAP at 887.5\,MHz (right). The coordinates for each itemised position listed in the WISE catalogue is marked with an ``*''.  The bottom right-hand corner shows the size and shape of the synthesised beam for each radio interferometer.}
    \label{fig_RCW38}
\end{figure*}

\begin{figure*}
\centering
\includegraphics[width=0.95\textwidth]{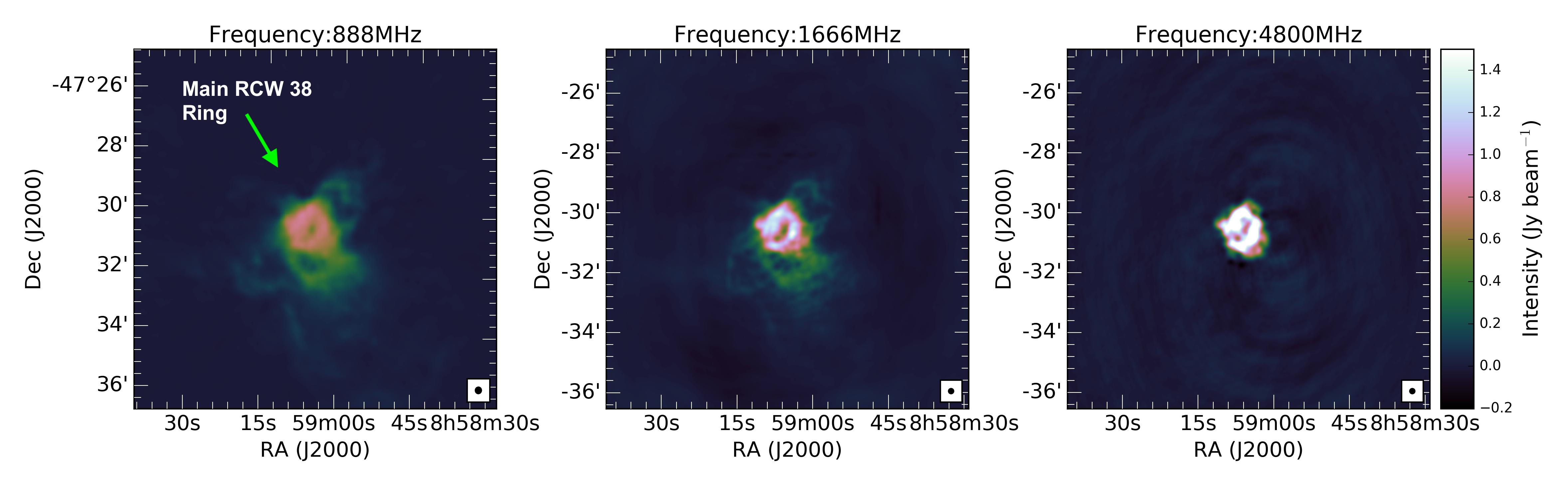}
\includegraphics[width=0.93\textwidth]{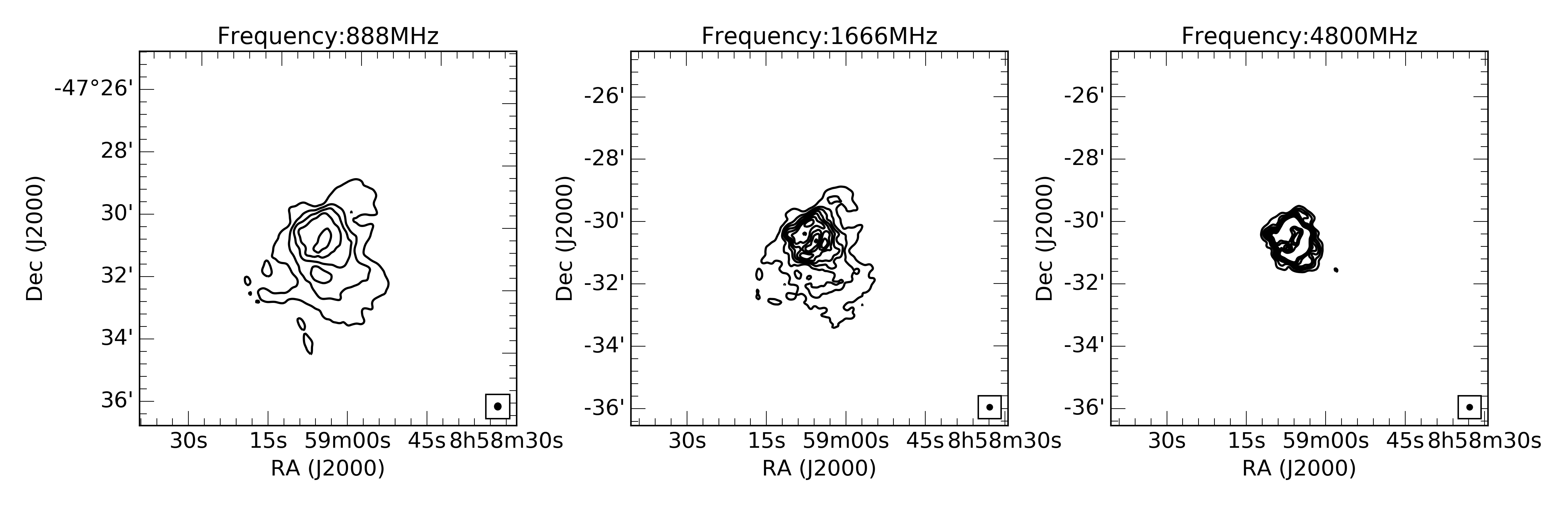}
\caption{Comparison of the emission morphological structure observed by ASKAP at 887.5\,MHz (left) by ATCA at 1666MHz (middle) and at 4800\,MHz (right). The synthesised beam for each image is shown on the bottom right of each image. The bottom row of images are the contour images for the same observations with contour levels set at 0.1--1.3\,Jy with an increment of 0.2\,Jy. The bottom right-hand corner shows the size and shape of the synthesised beam for each radio interferometer.}
\label{ATCA_ASKAP}
\end{figure*}

The area around RCW 38 at 114\,MHz from the MWA and 887\,MHz from ASKAP is shown in Figure \ref{fig_RCW38}. There is broad agreement between the 114 and 887\,MHz images, keeping in mind the 114\,MHz is of lower resolution and does not show the details afforded by the higher resolution of the 887\,MHz image. As shown in Figure \ref{ATCA_ASKAP}, there is also agreement for the IRS1 and IRS2 region between ASKAP at 887\,MHz and ATCA at 1666 and 4800\,MHz (the latter from \cite{Wolk_2008}, their Figure\ 5). When comparing the two lower frequencies of 114\,MHz and  887\,MHz, two interesting features emerge. One is that the 114\,MHz emission peaks to the South South-West of RCW 38, whereas the 887 MHz emission peaks at RCW 38, in the same ring-like structure centred on IRS2 shown in Figure \ref{ATCA_ASKAP}.

\begin{figure*}
    \includegraphics[width=0.85\textwidth]{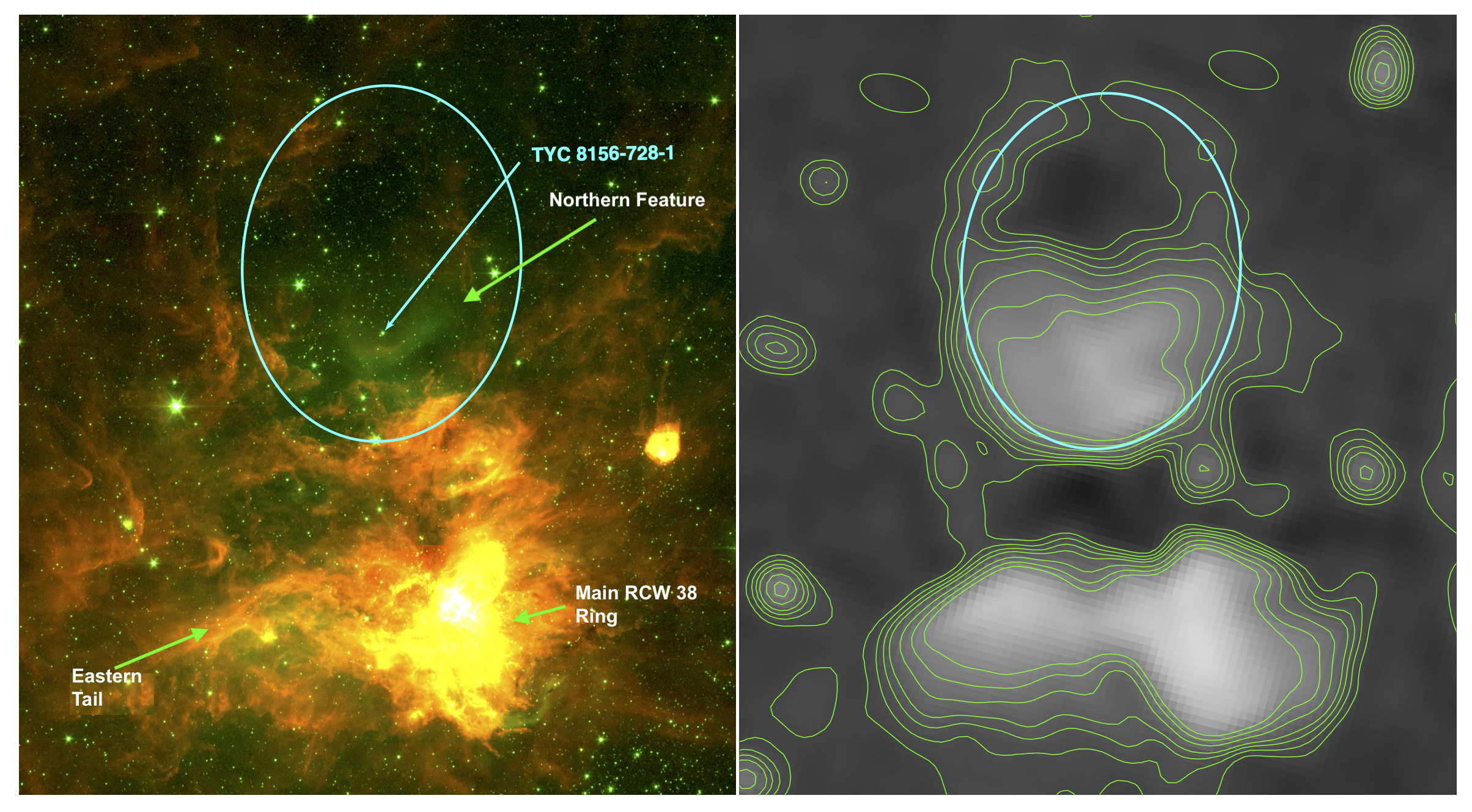}
    \caption{Bubble surrounding the O-star candidate TYC 8156-720-1. The right-hand image is the same field-of-view from the MWA 114\,MHz data on the left in Figure \ref{fig_RCW38} centred at RA 8:59:26, Dec --47:26:12. On the left is a Spitzer IRAC 3.6, 4.5, and 8\,$\mu$m three-coloured (blue, green, and red respectively) image of the region to the north of RCW 38 (visible at lower centre).  The oval shows the location of the proposed bubble due to TYC 8156, whose position is indicated on the Spitzer image.}
    \label{RCW38_Infrared_Low}
\end{figure*}

\begin{figure*}
    \includegraphics[width=0.98\textwidth]{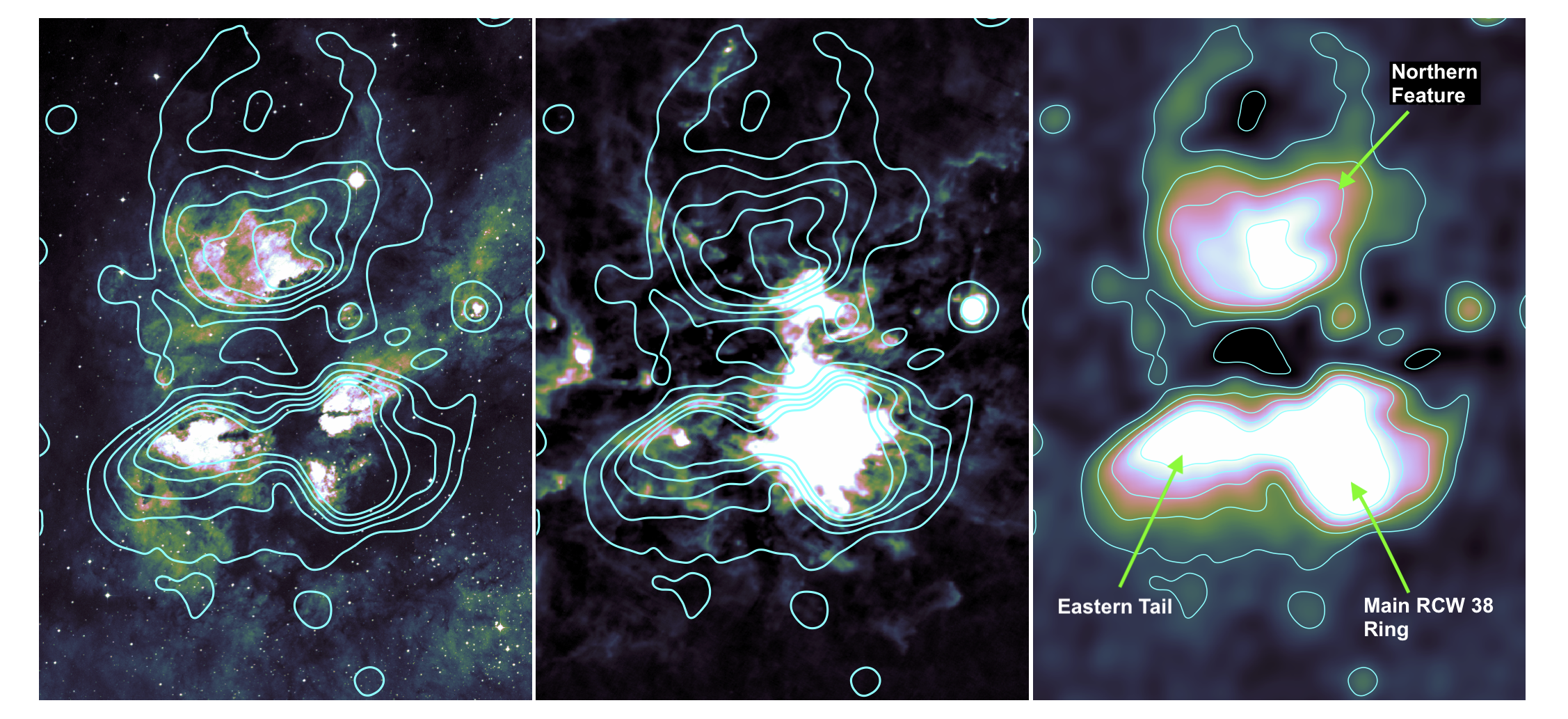}
    \caption{Image showing the radio emission at 114\,MHz (right; similar to Figure \ref{fig_RCW38} centred at RA 8:59:26, Dec -47:26:12, J2000) of RCW 38 overlaid onto the high-resolution H$\alpha$ optical emission from SuperCOSMOS H-alpha Survey (left) and the Herschel 165\,$\mu$m emission (middle). The contours span from 0 to 25\,Jy at six equal increments.}
    \label{Dust_Gas_RCW38}
\end{figure*}

Second is that the emission at 114\,MHz above the RCW 38 main lobe shows a clear asymmetric ring-like structure of its own, limb brightened in the south edge, which is barely visible at 887\,MHz (even with an intensity stretch).  This asymmetric ring shows good correspondence with IRAC band 2 emission at 4.8\,$\mu$m, band 4 emission at 8.0\,$\mu$m and the WISE 3.4 and 4.6\,$\mu$m (likely from shocks and PAHs, respectively), and seems to be associated with a candidate O-star TYC 8156-728-1 (Figure \ref{RCW38_Infrared_Low}; see also Figure~\ref{Dust_Gas_RCW38}; \citealt{Winston_2011}). The 4.6$\mu$m emission lies inside the South side of the ring, and corresponds to the region of bright 114\,MHz emission on the South side of the ring. Not much is known about this feature or the O-star candidate, further investigations are needed with high-resolution radio telescopes at a range of frequencies.

The tail toward the East of RCW 38, not clearly detected in either the 1666\,MHz or 4800\,MHz images, has faint emission in WISE 4.6\,$\mu$m, 887\,MHz continuum and 165\,$\mu$m emission and is strongest in the 114\,MHz and H$\alpha$. \cite{Urquhart_2007} observed $^{13}$CO 1--0 emission in this region and \cite{Broos_2013} discovered a young high-mass star candidate (2MASS J08595792-4732147) in the region using x-ray and infrared data. It may be possible that this region is in the later stages of development where the thick ionising dust cloud is being swept away by the stellar winds.  It is also possible that a single O star will have a less dense H{\sc ii} region as they form at a faster rate than B stars; thus not having the time to build up and ionise a thick cloud. However, as little is known about this region, future surveys and additional information will allow us to better interpret the data for this region.

The high resolution ($\sim$10'') radio maps of RCW 38 at 1666\,MHz (this work; see also \cite{Bourke_2004}) and 4800\,MHz \citep{Wolk_2008} obtained with the ATCA resolve the central region and show the main RCW 38 ring-like structure as seen at mid-infrared wavelengths, marking the interface of the H{\sc ii} region and the surrounding dense gas.  These data are shown in Figure \ref{ATCA_ASKAP}, along with the 887.5 MHz image from ASKAP, in which the ring is also evident.  The relationship of the 165\,$\mu$m emission in Figure \ref{Dust_Gas_RCW38} and the WISE 22\,$\mu$m emission in Figure \ref{WISE_ASKAP}, suggests that the continuum emission in the larger region called RCW 38 (the southern feature, eastern tail and the main lobe) is a mix of thermal and synchrotron radiation \citep{Tabatabaei_2007_B,Calzetti_2007,Bendo_2008}. \cite{Masque_2020} discuss that thermal emission is often dominant in the densest regions and the emission of non-thermal synchrotron radiation is dependant the presence of magnetic fields and shocks.

\begin{figure}
    \centering
    \includegraphics[width=0.48\textwidth]{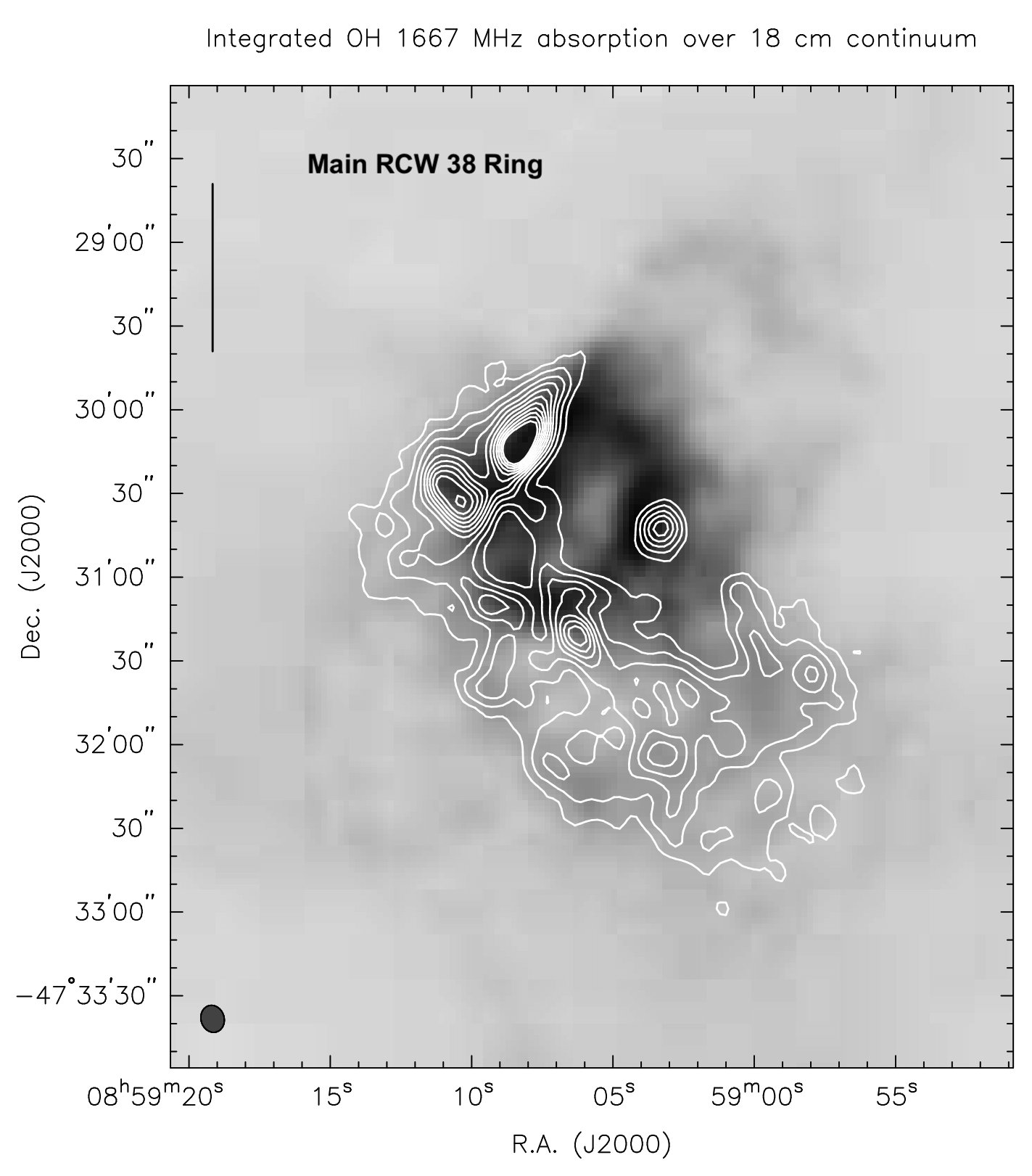}
    \caption{Image of integrated 1667 MHz OH absorption (contours) over the 1666\,MHz (18 cm) continuum (greyscale) observed with ATCA. The OH contours on this figure are from -0.65 to -7.15 in steps of 0.65 Jy km s$^{-1}$/beam. The greyscale ranges from -0.2 to 1.2 Jy/beam.  Velocity of integration is $-$4.5 to $+$7.5\,km\,s$^{-1}$. The beam is shown at lower left, and the vertical bar indicates 0.5 pc at the distance of RCW 38.}
\label{Molecules_Overlay}
\end{figure}

The integrated main line OH 1666\,MHz absorption is shown in Figure \ref{Molecules_Overlay}. No OH masers are present.  While the absorption ``peaks'' (i.e., troughs) fall on top of the radio continuum ring, there is extended absorption to the south, where the 114\,MHz continuum peaks (and where the radio optical depth is modest). A comparison of the Parkes data published in \cite{Bourke_2001} and ATCA spectral line data indicates that the component giving rise to the Parkes Zeeman detection is probably a compact, isolated absorption feature located on the IRS1 ridge. The interface between the radio continuum ring and the IRS1 dense gas is a photodissociation region, where the stellar winds and ultra-violet radiation from IRS2 are impacting the cloud, perhaps compressing it and causing triggered star-formation.  Additional observations to study this region in all four of the OH transitions, including Zeeman splitting will be undertaken shortly with ATCA, to better understand the role the magnetic field is playing in the star formation within IRS1.  A comparison of the integrated OH absorption spectrum with the integrated $^{13}$CO 1-0 spectrum \citep{Bourke_1999} shows their line shapes are very similar.  As these two transitions are expected to be co-located, tracing similar material, this suggests that the  $^{13}$CO can be used to infer the physical conditions of the OH gas. This will be explored in a future study.  

\subsection{Radio Spectral Index}
Although the $u,v$ coverage of the 1666\,MHz data is more sparse than that of the 887.5\,MHz image\footnote{ATCA has six antennas, and for our observations, the baselines ranged between 105\,m and 6\,km and ASKAP has 36 antennas, with baselines between 22\,m and 6\,km.}, particularly at short $u,v$ distances, such that some of the more extended emission is missing from the 1666\,MHz image, we have constructed a two-point spectral index map between 887.5 and 1666\,MHz. We have not added in the MWA data at 114\,MHz as the resolution of $\sim$1\,arcmin does not allow for the same detail within the region as the ASKAP and ATCA data.  Both the ATCA \& ASKAP images have been convolved to the same resolution  of 15\,arc\,sec. At these frequencies we expect any H{\sc ii} region to be primarily in the optically thick regime, or at least in the turn-over from optically thin at higher frequencies (spectral index $-$0.1) to thick at lower frequencies (spectral index 2.0). The spectral index should therefore be mostly positive throughout the region but not greater than 2, assuming the region is well resolved at both wavelengths.  

\begin{figure}
    \centering
    \includegraphics[width=0.48\textwidth]{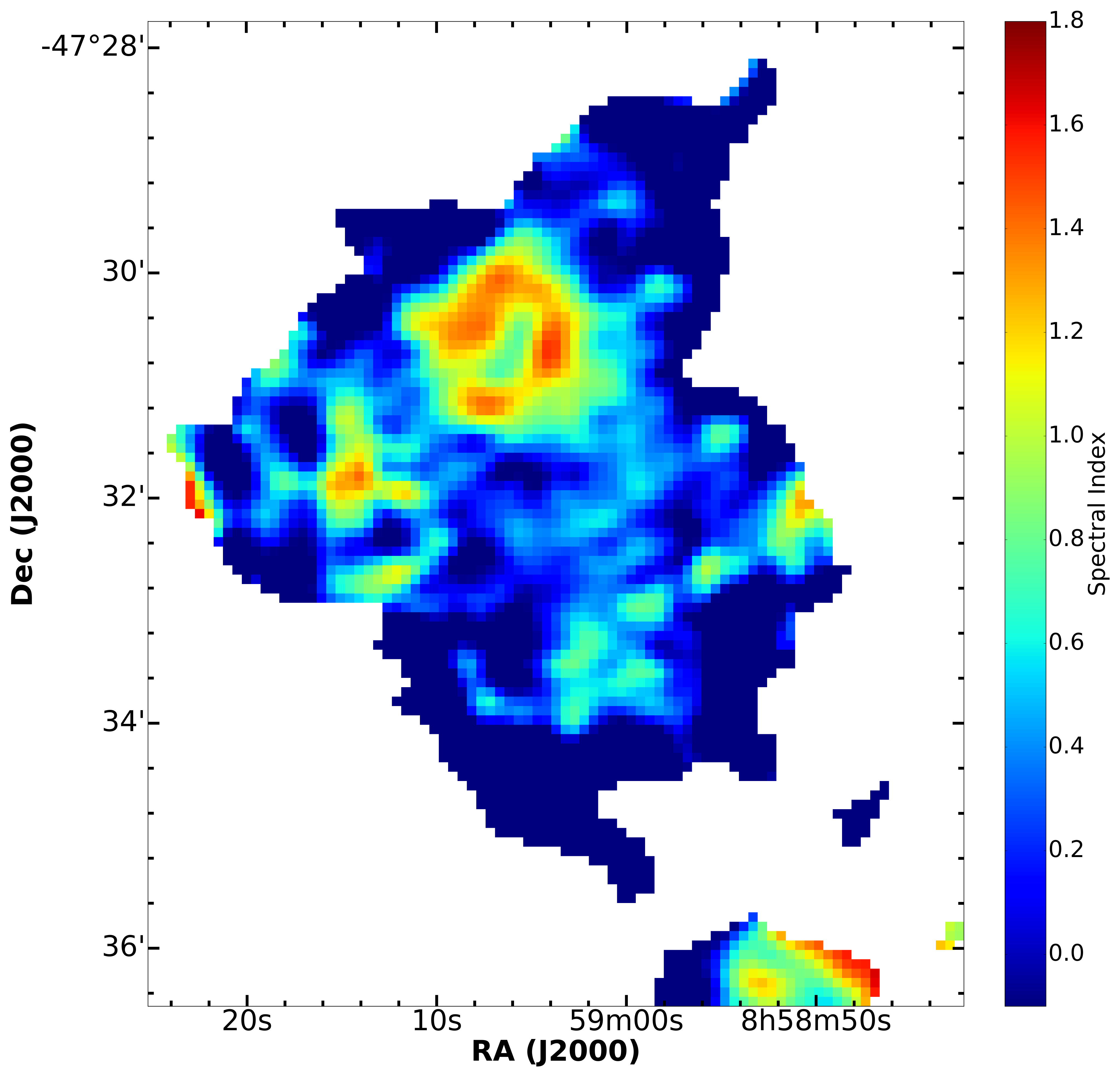}
    \caption{Radio spectral index map of RCW 38 using ATCA 1666\,MHz and ASKAP 887.5\,MHz data.}
\label{SED}
\end{figure}

The spectral index map shown in Figure \ref{SED} supports these assumptions, with the median optical depth value of 0.2$\pm$0.07 representing most of the map, peaking in the ring around IRS2 with a value of 1.6$\pm$0.3. The errors here are generated from a root mean square error map of the two input maps. Although we note this may overestimate the errors \citep{Dickey_2021}, it demonstrates that the different segments of the region can be correctly differentiated. This map illustrates the potential value in constructing similar maps of candidate H{\sc ii} regions in a larger survey with matching $u,v$ coverage using frequencies at the low, mid, and high end of the ASKAP band (upcoming RACS survey full data release), to study and confirm the nature of H{\sc ii} regions.  Additionally with these upcoming wide-field, high-resolution surveys, there will be significant improvements in our understanding of the optical depths and relationships to the spectral index and electron densities \citep{Dickey_2021}.

\section{Other Prominent Sources in the Field}
Within the Vela region, the Gum Catalogue \footnote{http://galaxymap.org/cat/list/gum/1} and RCW Catalogue \citep{RCW_Catalog} provide the identification of several H{\sc ii} regions observable in the southern hemisphere. Other candidate and known H{\sc ii} regions are also collected in the WISE H{\sc ii} Catalogue \citep{Anderson_2019}, in \cite{Wenger_2021}, and in {\sc SIMBAD}. In the foreground, at $\approx$300-450\,pc away is the Gum Nebula, the Vela Supernova Remnant, and prominent high-mass stars.  In the background are a series of H{\sc ii} regions associated with a giant molecular cloud along the Galactic Plane, at a distance of 1.7 to 1.8\,kpc away and some sources are thought to be further in the background.  These are summarised in Table \ref{Other_S}. 

Overall, the field includes a complex structure of foreground and background sources that are difficult to disentangle.  Below are details of new, wide-field observations of H{\sc ii} regions observed with ASKAP and the MWA which provide greater detail around the clusters containing H{\sc ii} regions.  Any H{\sc ii} regions that were analysed and discussed in \cite{Hindson_2016}, except for RCW39, are not included in this survey as it would be a repetition of information. RCW39 is included here as the ASKAP image provides greater detail of the source structure than previously observed in other surveys.

\begin{table*}
\small
\caption{Characterisation of H{\sc ii} regions in the field covered in this paper.  The distances listed in column 3 are from various sources and are sometimes assumed in literature, as specified as ``M" for measured and ``A" for assumed. The values in columns 4--7 with an ``*" denote measurements determined within this work.  The total luminosity (Log Lum.) and Lyman Continuum Photon Flux (N$_c$) are calculated based on the equations and assumption listed in Appendix A. The luminosity values calculated from IRAS fluxes are an upper limit as the beam for IRAS was large. Ionised gas velocity measurements are derived from hydrogen radio recombination lines, except where noted. All information from previous work is referenced in each section for the individual sources and summarised here.}
\label{obs}
\raggedright
\begin{tabular}{lccccccc}
\hline
Source	&WISE Name	&Distance	&	Stellar Pop.	&	Age	&	Log (Nc)	&	Log (Lum.)&Velocity 	\\
&	& (kpc) & & (Myr) & (photons\,s$^{-1}$) & (L$_{\odot}$)&(km\,s$^{-1}$)\\
	\hline
	\hline
RCW38 (Main Ring)& G267.935--01.075	&	1.63$\pm$0.17	&	40 OB Stars	&	<1	&	49.4	&	6.09& 1.00$\pm$0.1$^{a}$	\\
IRAS08563-4711& G267.730--01.09&2.1& At least 1 B1$^{**}$ & &45.63* & 3.66& 6.8$\pm$0.3$^{a}$\\
RCW39	& G269.068--01.114&	2.6	& At least 1 O9.5$^{**}$		&		&	45.9*	&4.6$^{*}$&		13.5$\pm$0.1$^{a}$\\
RCW40	& G269.174--01.436 &	1.7	&O8		&	Indeter.	&	48.8	&	3.5*&7.5$\pm$2.1$^{a}$	\\
RCW42	& G274.004--01.152&	6.4	&	>12 OStars	&Late Stage		&	50.30	&	Confused&39.5$\pm$0.1$^{a}$	\\
GAL268.4--0.9	& G268.419--0.843&	1.7$-$1.8	&	At least 1 B0 Star	&	0.8	&	47.0$^*$	&	5.0* & 5.4$\pm$0.1$^{a}$	\\
IRAS09015-4843& G269.454--01.468 & 7.80$\pm$0.40 &At least 1O9.5$^{**}$ &&47.7*&2.78& 67.0$\pm$0.3$^{a}$\\
GAL263.62--00.53& G263.615--0.534 &0.7&B1&1--2&ND&3.0*&11.0-17.5$^{b}$\\
RCW41 &G270.255+00.85& 1.3 $\pm$ 0.2& O9V star& 2.5--5&45.5*&3.7*&2.5-5.5$^{b}$\\
GAL281.01--01.53&G281.047--01.541 &1.9& &&ND&ND&--5.6$\pm$0.3$^{a}$\\

\hline
\end{tabular}\\
\label{Other_S}
*Determined as part of this work.  All other values are determined in other publications as referenced in the text.\\
**Based on Table 1 in \cite{Panagia_1973}\\
ND=Not Determined\\
RRL Velocity Measurements: $^{a}$ \cite{Wenger_2021} Table 10 (H88-H112)\\
$^{b}$Methanol maser peak velocity measurements associated with the complex from \cite{Green_2012}, which can be used as a proxy for the systemic velocity to within $\sim$3\,km/s \citep{Green_2011}.\\
\end{table*}

\subsection{IRAS08563-4711}

The candidate H{\sc ii} region IRAS08562-4711 is one arc minute in RA from the main lobe of RCW 38 at a position of RA 08:58:04.2, Dec -47:22:57 (J2000).  The $Gaia$ distance catalogue lists the photometric distance for stellar candidates in the region as ranging from 1.3 to 2.5\,kpc and the $^{18}$CO observations from \cite{Torrii_2021} suggests there may be some interaction with RCW 38. 

The source is unresolved with both ASKAP and the MWA, as shown in Figure \ref{fig_RCW38} and \ref{WISE_ASKAP}. \cite{Navarete_2015} observed the source with MSX for H$_{2}$ (source 217 in their paper) and classified it as being two diffuse or filamentary high-mass star clouds containing a cluster of embedded stars. They determined a kinematic distance of 2.1\,kpc and luminosity of 10$^{3.66}$\,L$_{\odot}$. \cite{Valdettaro_2007} searched the region for water masers, a powerful tracer of embedded stars, but did not detect any with the Tidbinbilla 70\,m dish.  Using the flux density from the ASKAP observations and the distance of 2.1\,kpc, we determined the the stellar H{\sc ii} region to contains at least one B1 star from using \cite{Panagia_1973} and the ionising photon flux listed in Table \ref{Other_S}.  

As shown in Figure \ref{Dust_Gas_RCW38}, the region has both strong H$\alpha$ and 165\,$\mu$m emission.  This suggests it is a compact region of hot, dense ionising gas consistent with an H{\sc ii} region.

\subsection{RCW 39}
\begin{figure*}
\centering
\includegraphics[width=0.9\textwidth]{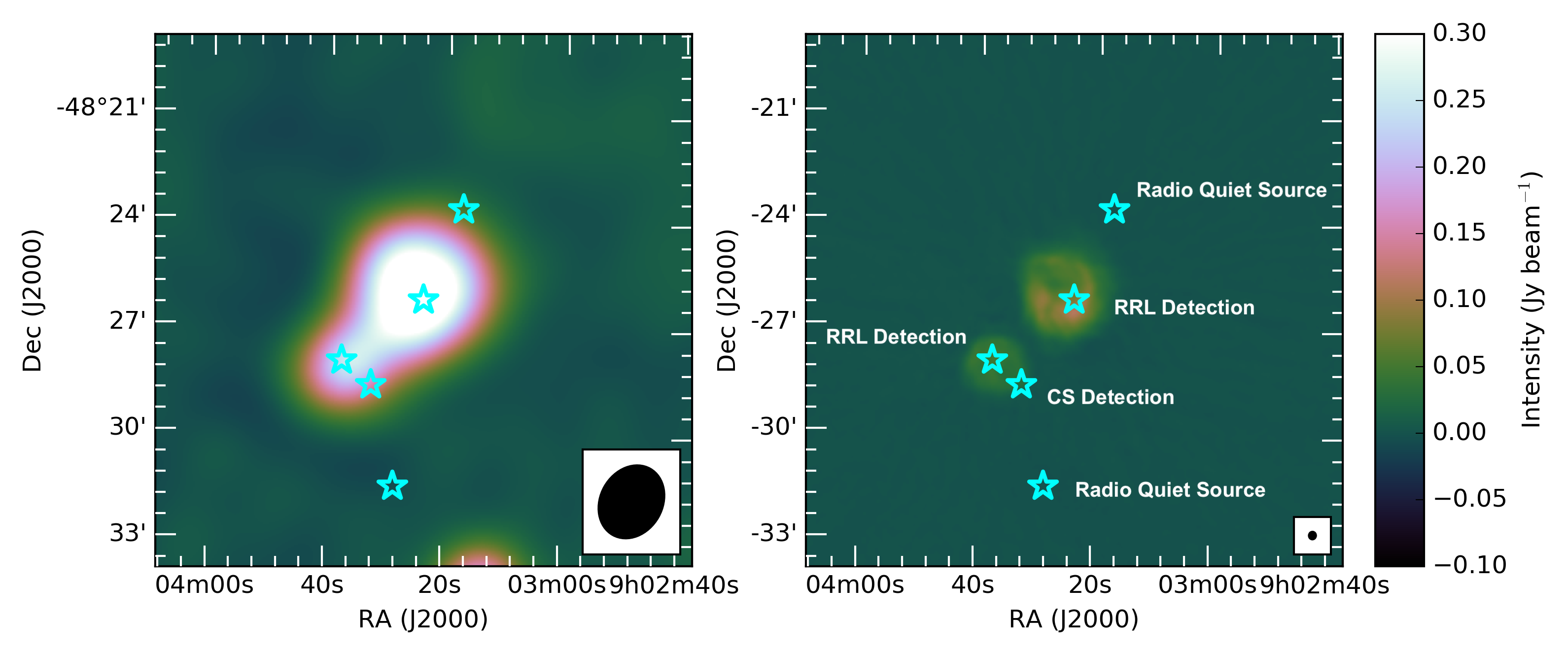}
\caption{RCW 39 (GAL269 -1.14) HII region imaged in continuum with MWA at 114\,MHz (left) and ASKAP at 887.5\,MHz (right).  The position of the ``*'' marks are from the WISE Catalog of sources.  The bottom right-hand corner shows the size and shape of the synthesised beam for each radio interferometer.} 
\label{Gal269}
\end{figure*}

The GAL269.13--1.14 H{\sc ii} region within the cluster RCW 39, at the position of RA 09:03:23, Dec --48:26:34 (J2000), is comprised of two ring-like clouds as shown in Figure \ref{Gal269}. This source was included in a census of H{\sc ii} regions by \cite{Caswell_1987} with the Parkes Radio Telescope at 5\,GHz.  They determined a continuum flux density of 6.8\,Jy for the source and velocities of 10 and 15\,km\,s$^{-1}$ for the H$_{2}$CO and Hydrogen recombination lines respectively.  \cite{Wenger_2021} measured the velocity for the Eastern lobe as 14.3\,km\,s$^{-1}$ and 12\,km\,s$^{-1}$ for the south-western region.  \cite{Bronfman_1996} measured CS(2-1) transition with the Swedish-ESO Submillimetre Telescope with a velocity of 10.3\,km\,s$^{-1}$.  \cite{Hou_2014} report the region to be at the distance of 2.6\,kpc in their catalogue via the method of kinematic distance using a Galaxy rotation curve with the current IAU standard, but within the WISE Catalogue it is listed at a distance of 3.09$\pm$0.62\,kpc.

In our observations, the ASKAP data separates the region into two components where previous studies have shown one blurred region similar to what is observed with the MWA (Figure \ref{Gal269}).  This source was also observed by \cite{Hindson_2016} with the MWA from 70--200\,MHz finding a spectral index at these low frequencies of 2.2$\pm$0.4.   From the observations so far reported in this region, it is unknown if these two clouds are interacting.

Using the flux density from the ASKAP observations and the distance of 2.6\,kpc, we determined the the stellar H{\sc ii} region located in the north-east contains at least one O9.5 star from using \cite{Panagia_1973}. By using the equations for the total luminosity from IRAS fluxes and the continuum photon flux (N$_c$) in Appendix A, we characterise this lesser known region.  We put an upper limit for the total luminosity as 10$^{4.6}$\,L$_{\odot}$. These values are collected in Table \ref{obs}.

\subsection{RCW40}

RCW40 (within the Gum 25 cluster) is a bright southern H{\sc ii} region characterised as containing a cluster of hot, young, massive stars at the position RA 09:02:21.3, Dec --48:41:55 (J2000).  \cite{Copetti_2000} measured the distance to be 1.7\,kpc with integrated photometry which is updated from the distance of 1.4\,kpc reported by \cite{Shaver_1983}.  \cite{Reyle_2002} found the cluster to have a large dispersive colour in infrared, suggesting a strongly varying extinction exists in front of or within the cluster.  This means that they could not reliably determine the age of the cluster.


\cite{Reyle_2002} obtained an average visual extinction (A$_{V}$) of 9 mag toward the cluster using the Deep Near Infrared Survey of the Southern Sky (DENIS), and so it contains an embedded young cluster. This is a more sensitive measurement over the 2MASS catalogue values \citep{Cutri_2003}. \cite{Kennicutt_2000} studied H{\sc ii} regions within the Milky Way, Small Magellanic Clouds, and Large Magellanic Clouds.  Using optical data they calculated the stellar temperature (T$_{*}$) of 38,400\,K and the ionising luminosity of the stars of 10$^{48.80}$\,photon\,s$^{-1}$. \cite{Hawley_2011} reported a comparison of O{\sc ii},  O{\sc iii}, and N{\sc ii} abundances for the region.  In a survey of H{\sc ii} regions by \cite{Goy_1980}, they classified the region as being dominated by an O8 spectral type high-mass star.

\begin{figure*}
\centering
\includegraphics[width=0.9\textwidth]{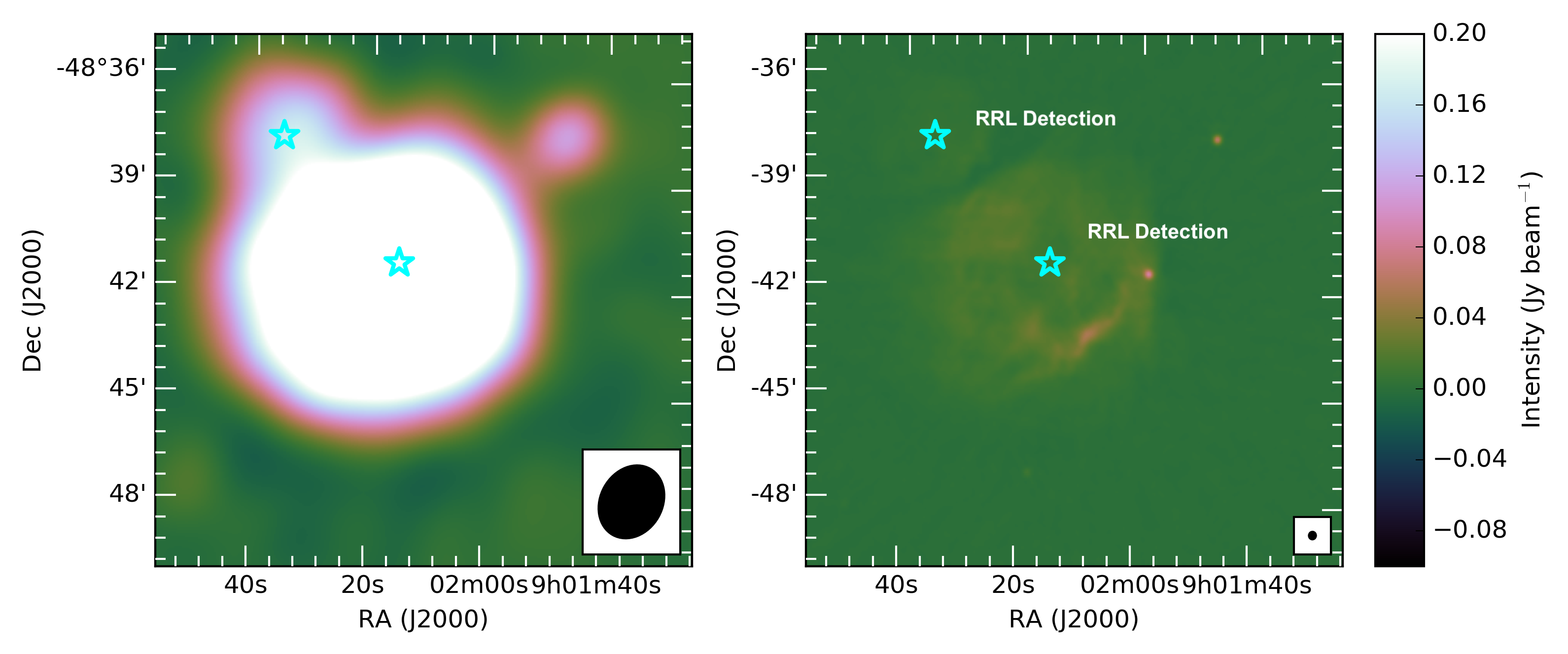}
\caption{RCW40 (Gum 25 Cluster) imaged in continuum with MWA at 114\,MHz (left) and ASKAP at 887.5\,MHz (right).  The ``*'' on each image marks the positions of RRL detections. The bottom right-hand corner shows the size and shape of the synthesised beam for each radio interferometer.}
\label{Gum25}
\end{figure*}

\cite{Wenger_2021} measured recombination lines at the two locations marked with an ``*'' on Figure \ref{Gum25}.  The southern lobe has a velocity of 3.8\,km\,s$^{-1}$ and the northern lobe has a velocity of 2.7\,km\,s$^{-1}$.  Previous to this work the focus of \cite{Kennicutt_2000} and other surveys is on the larger ring structure of the source.  

The continuum images from the MWA and ASKAP show a consistent structure for this source. The ASKAP image may suggest there is a separation between the two ``lobes'' of the H{\sc ii} region as seen in the optical images from the Digital Sky Survey (DSS).  Using the equations in Appendix A, we put an upper limit on the of the total infrared luminosity of 10$^{3.5}$\,L$_{\odot}$. 

\subsection{RCW42 (Gum 26 Cluster)}
RCW42 (Gum 26 Cluster) is a stellar cluster at RA 09:24:30.1, Dec --51:59:07 (J2000).  This Galactic H{\sc ii} region is part of a larger structure shell called GSH 277$+$0.36 at a distance of 6.5\,kpc \citep{McClure-Griffiths_2003}.  \cite{Moises_2011} studied RCW42 as part of a survey of Galactic H{\sc ii} regions in the near and far-infrared.  They found the region to be a densely-pact cluster of stars surrounded by a reddish nebula of about 600\,pc in diameter and with emission dominated at 8.0\,$\mu$m. 

\cite{Conti_2004_MSX} used the distance of 6.4\,kpc from \cite{Russeil_2003} to calculate the Lyman continuum luminosity of (N$_c$)=10$^{50.30}$.  From this \cite{Moises_2011} determined at least a dozen early O-type stars are associated with the region.  Although the region is complicated, \cite{Moises_2011} suggested that the kinematic distance and stellar sequence are correctly interpreted based on the near-infrared J, H and K three coloured image and Spitzer images \footnote{Courtesy of NASA/JPL-Caltech. http://www.spitzer.caltech.edu/}.  They also classify the region as a ``naked photosphere" and suggest it is quite late in the evolutionary phase. 

\begin{figure*}
\centering
\includegraphics[width=0.9\textwidth]{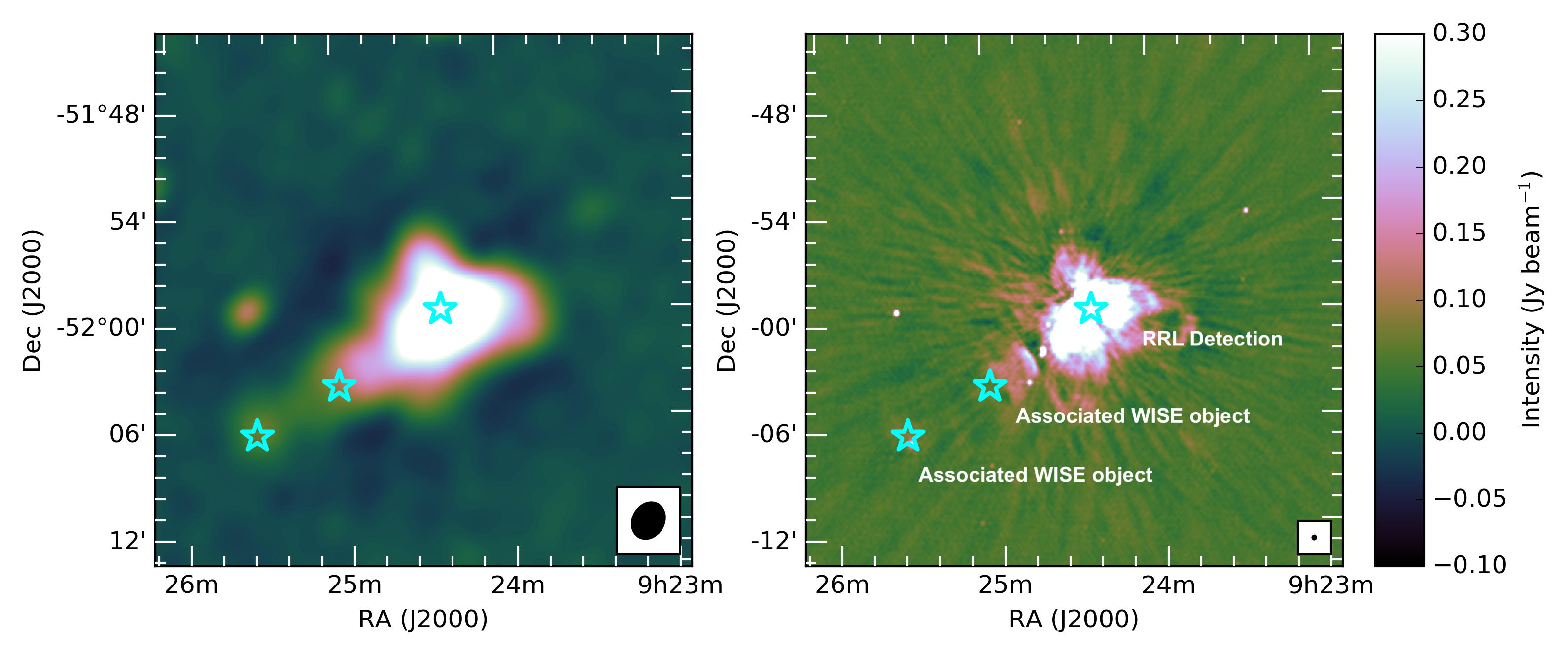}
\caption{RCW 42 H{\sc ii} region, part of the Gum 26 Cluster, imaged in continuum with MWA at 114\,MHz (left) and ASKAP at 887.5\,MHz (right).  There is a strong correlation with the MWA features with optical H$\alpha$. The bottom right-hand corner shows the size and shape of the synthesised beam for each radio interferometer.}
\label{Gum26}
\end{figure*}

The MWA continuum image mimics the structure detected in H$\alpha$ by the SHASSA survey \citep{Gaustad_SHASSA} and WISE 22\,$\mu$m suggesting a warmer gas is present. This is consistent with the idea the region is in the late stages of star formation and thus containing hotter dust emission. The ASKAP image shows greater structure detail.  The different points entries (marked with ``*'') from the WISE catalogue are on each image in Figure \ref{Gum26}.  The RRL detection by \cite{Anderson_2012} at a velocity of 39\,km\,s$^{-1}$, is in the central location of the low-frequency emission region. The source was included in the Class II methanol maser survey by \cite{Walsh_1997}, but no methanol was detected.

\subsection{GAL 268.4-00.9 H{\sc ii} Region}
The H{\sc ii} region GAL 228.4--00.9 (also known as IRAS 09002-4732) is located at RA 09:01:54.3, Dec --47:43:59 (J2000).  The parallax measurements from ESA’s Gaia satellite suggests a distance of 1.7--1.8\,kpc for the cluster \citep{Getman_2019} which places it along the edge of the Vela Molecular cloud, inline with RCW 38. \cite{Getman_2019} completed a survey of the little-known region using x-ray and infrared data. They found the cluster to be only about 0.8\,Myr old but twice as dense in stellar objects as the ONC.  They also classified it as a compact H{\sc ii} region within a filament of the Vela Molecular Cloud.

\cite{Lenzen_1991} studied the region and with an assumed distance of 1.8\,kpc calculated the Bolometric Luminosity of 10$^{5.0}$\,L$\odot$ and determined the region has at least one spectral type B0 star.  We use the Equation 1 and 3 from \cite{Kurtz_1994} and found a Lyman continuum photons per second ((N$_{c}$) to be 10$^{47.0}$\,photon\,s$^{-1}$ when making the same assumption as \cite{Lenzen_1991} of a distance of 1.8\,kpc, a T$_{e}$ of 10$^{4}$\,K and a dust to gas ratio of 100.  We use the flux density at 887.5\,MHz from the ASKAP data of 54.5\,mJy.

\begin{figure*}
\centering
\includegraphics[width=0.96\textwidth]{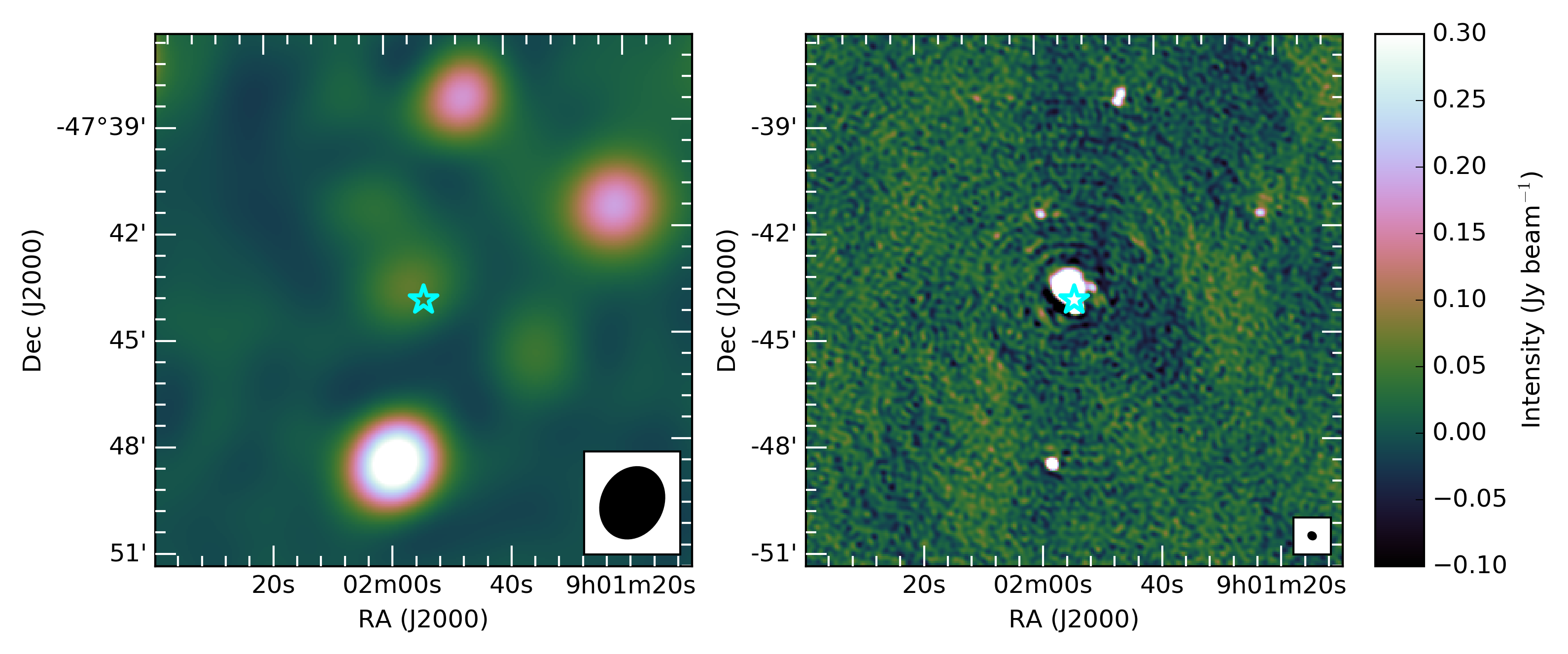}
\includegraphics[width=0.96\textwidth]{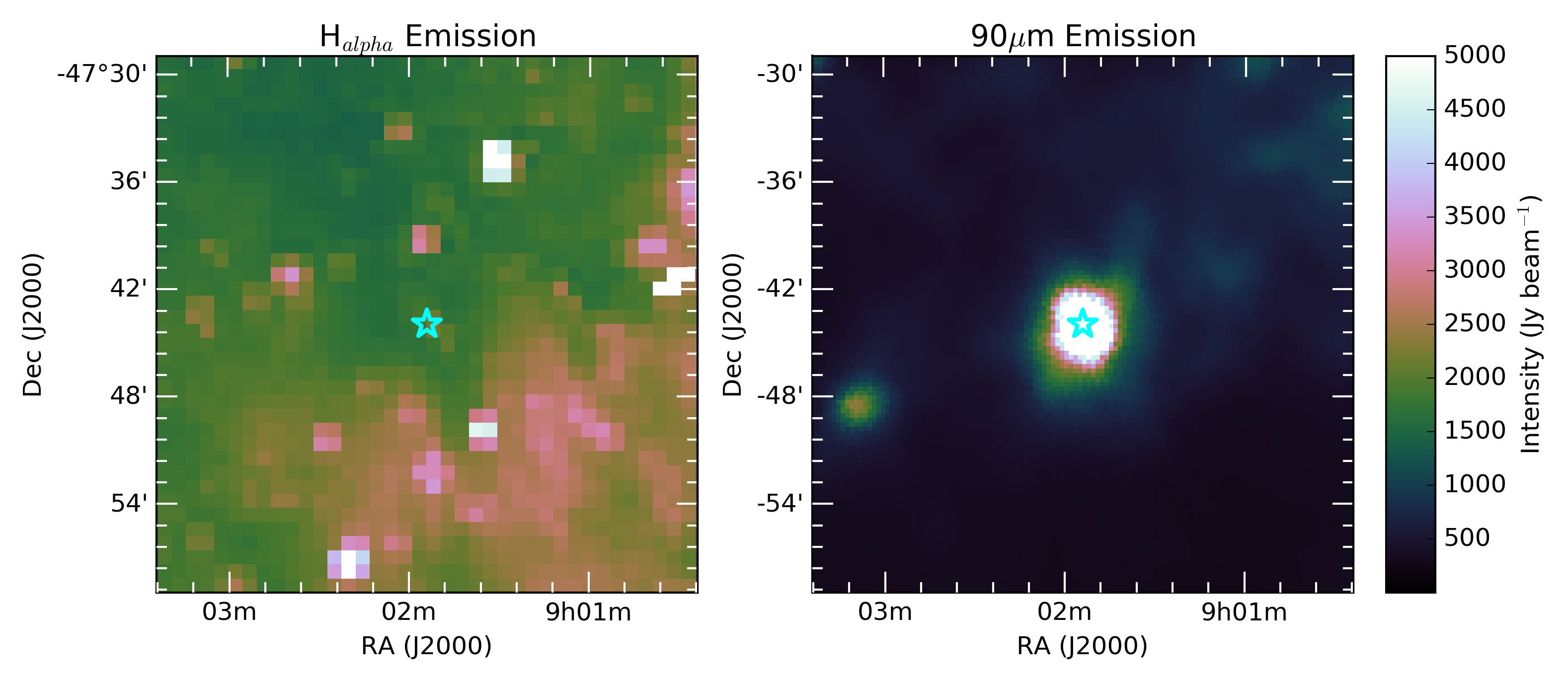}
\caption{Gal 268.4--00.9 H{\sc ii} imaged in continuum with the MWA at 114\,MHz (top left) and ASKAP  at 887.5\,MHz(top right). The bottom right image shows the dust emission at 90\,$\mu$m surveyed by AKARI. The bottom left image is optical H$\alpha$ emission from the Southern H-Alpha Sky Survey Atlas (SHASSA; \citealt{Finkbeiner_2003_Halpha}). The ``*'' is the position of the H{\sc ii} region sepcified in the WISE catalogue. The bottom right-hand corner shows the size and shape of the synthesised beam for each radio interferometer.}
\label{Dust_Halpha_Gal2684}
\end{figure*}

Figure \ref{Dust_Halpha_Gal2684} shows that the MWA image at 114\,MHz only contains a very faint emission toward the H{\sc ii} region.  There is strong emission toward the H{\sc ii} region with ASKAP at 887.5\,MHz as an unresolved source.  The 90\,$\mu$m dust emission map in Figure \ref{Dust_Halpha_Gal2684} shows a matched structure with ASKAP toward the H{\sc ii} region but little H$\alpha$ emission.  This suggests the region is dominated by cold dense gas, which is consistent with the analysis done by \cite{Getman_2019} that the H{\sc ii} region is young.  The ``*'' on the images marks the location of the H{\sc ii} region as listed in the WISE Catalogue. The RRL detection by \cite{Anderson_2012} is at a velocity of 5\,km\,s$^{-1}$.

\subsection{IRAS09015-4843}
\begin{figure*}
\centering
\includegraphics[width=0.96\textwidth]{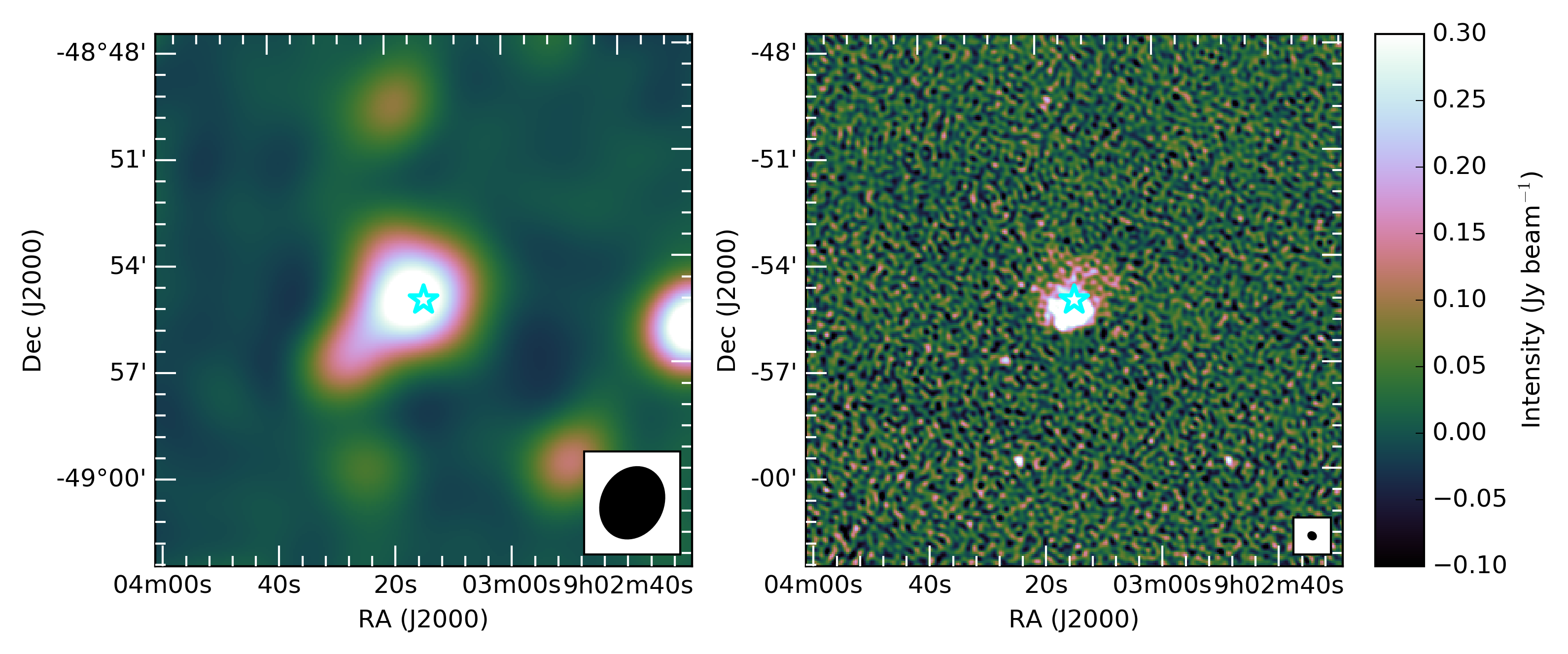}
\caption{IRAS09015-4843 H{\sc ii} imaged in continuum with the MWA (left) at 114\,MHz and ASKAP (right) at 887.5\,MHz.  The ``*'' symbol marks the position of the ionisation source as listed in the WISE catalog. The bottom right-hand corner shows the size and shape of the synthesised beam for each radio interferometer.}
\label{IRAS09015}
\end{figure*}

The compact H{\sc ii} region known as IRAS09015-4843 is located at RA 09:03:13.5, Dec --48:55:21 (J2000). The distance is listed in the WISE H{\sc ii} catalogue as 7.80$\pm$0.40\,kpc \citep{Anderson_2015_Distance}.  \cite{Walsh_1997} observed Class II methanol masers (CH$_3$OH) toward the region and \cite{Breen_2011} found water masers, both signs of active star formation.  \cite{Anderson_2012} observed hydrogen RRLs at a velocity of 71\,km\,s$^{-1}$.  \cite{Green_2012} measured the velocity of 6.7\,GHz methanol masers for the region between 53.5 to 56.6\,km\,s$^{-1}$. Observations of $^{12}$CO \citep{Wouterloot_1989} were also made toward the ionisation centre labelled by an ``*'' on Figure \ref{IRAS09015}. Both the ASKAP and MWA images shows a non-resolved point source.  

\cite{Hill_2009} completed an analysis of Galactic H{\sc ii} regions through the infrared spectral index. They found the cold gas temperature between 2.7--35\,K, the clump mass to be at least 3.1$\times$10$^{2}$\,M$_{\odot}$ and a luminosity of at least 6.1$\times$10$^{2}$\,L$_{\odot}$. Using the continuum emission from RACS at 887\,MHz, we determine the continuum photon flux $N_c = 10^{47.7}$\,photons\,s$^{-1}$.  

\subsection{GAL263.62-00.53}
The H{\sc ii} region GAL263.62-00.53 is in the embedded cluster IRAS08438-4340 (IRS16) \citep{Massi_2010} and one of six clusters noted within Vela Molecular Ridge Cloud D \citep{Murphy_1991}. The position of the region is at RA 08:45:34.9, Dec, --43:51:07 (J2000). \cite{Massi_2006} studied the region and by using the initial mass function found the clusters in Vela D to be of similar age which was constrained to 1--5\,MYr and IRS16 had a total cloud mass of 99\,M$_{\odot}$.  \cite{Massi_2010} further studied the region in detail using near and far infrared analysis.  Using two different methodologies they constrained the age of the cluster to 1--2\,MYr and found the H{\sc ii} region was fuelled by a Early B star. 

\cite{Caswell_1987} constrained the size of the radio emission to one arc\,second at a distance of 700\,pc. \cite{Massi_2010} suggested that this means that the H{\sc ii} region would be greater than the Str{\"o}mgren sphere only for an ionising star later than B1, which was consistent with previous results by \cite{Massi_2003} which suggested a spectral type B0-B1 star fuelled the H{\sc ii} region.

\begin{figure}
\centering
\includegraphics[width=0.48\textwidth]{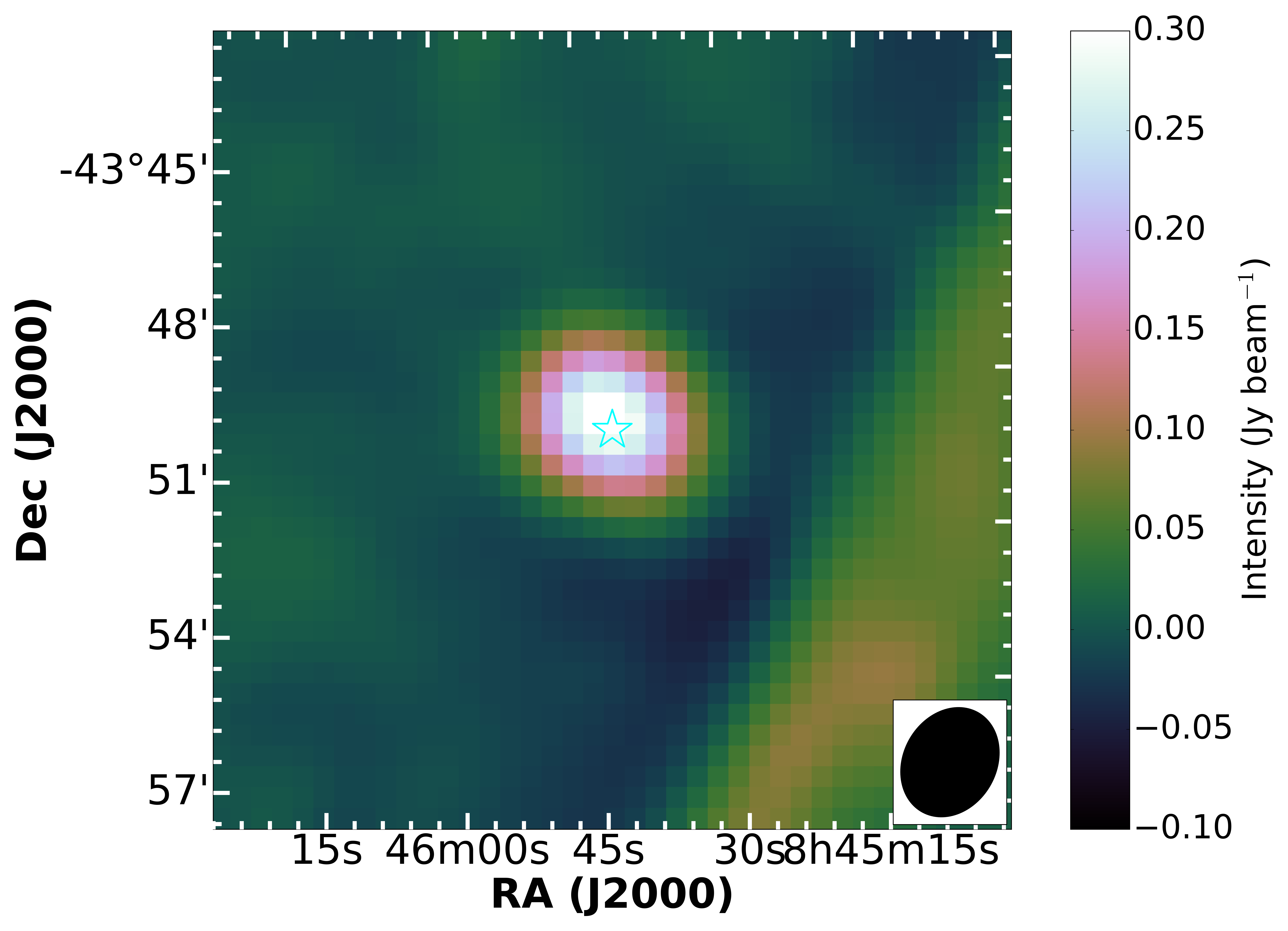}
\caption{GAL263.62-00.53 H{\sc ii} imaged in continuum with the MWA at 114\,MHz.  
The ``*'' symbol marks the position of the ionisation source as listed in the WISE catalog. The bottom right-hand corner shows the size and shape of the synthesised beam for each radio interferometer.}
\label{GAL263}
\end{figure}

\cite{Massi_2018} studied the millimetre continuum emission of the region and found 3 dense cores around the GAL263.62-00.53 region.  They suggest that the expanding ionisation cloud is spurring further star formation in the area. As shown in Figure \ref{GAL263}, GAL263.62-00.53 is unresolved in the MWA image.  The ``*'' on the image marks the location of the RRL detection by \cite{Wenger_2021}, which had a velocity of -0.2\,km\,s$^{-1}$.  \cite{Green_2012} measured the 6.7\,GHz methanol maser velocity to range from 11.0 to 17.5\,km\,s$^{-1}$ for the region. This source was between two footprints from the RACS survey, so a high resolution image was not obtained.  This also meant we could not obtain a good measure of the photon flux for the region.  Using the IRAS fluxes and the equations in Appendix A, we determined the upper limit of the luminosity of 10$^{3.0}$\,L$_{\odot}$.

\subsection{RCW41}
\begin{figure*}
\centering
\includegraphics[width=0.96\textwidth]{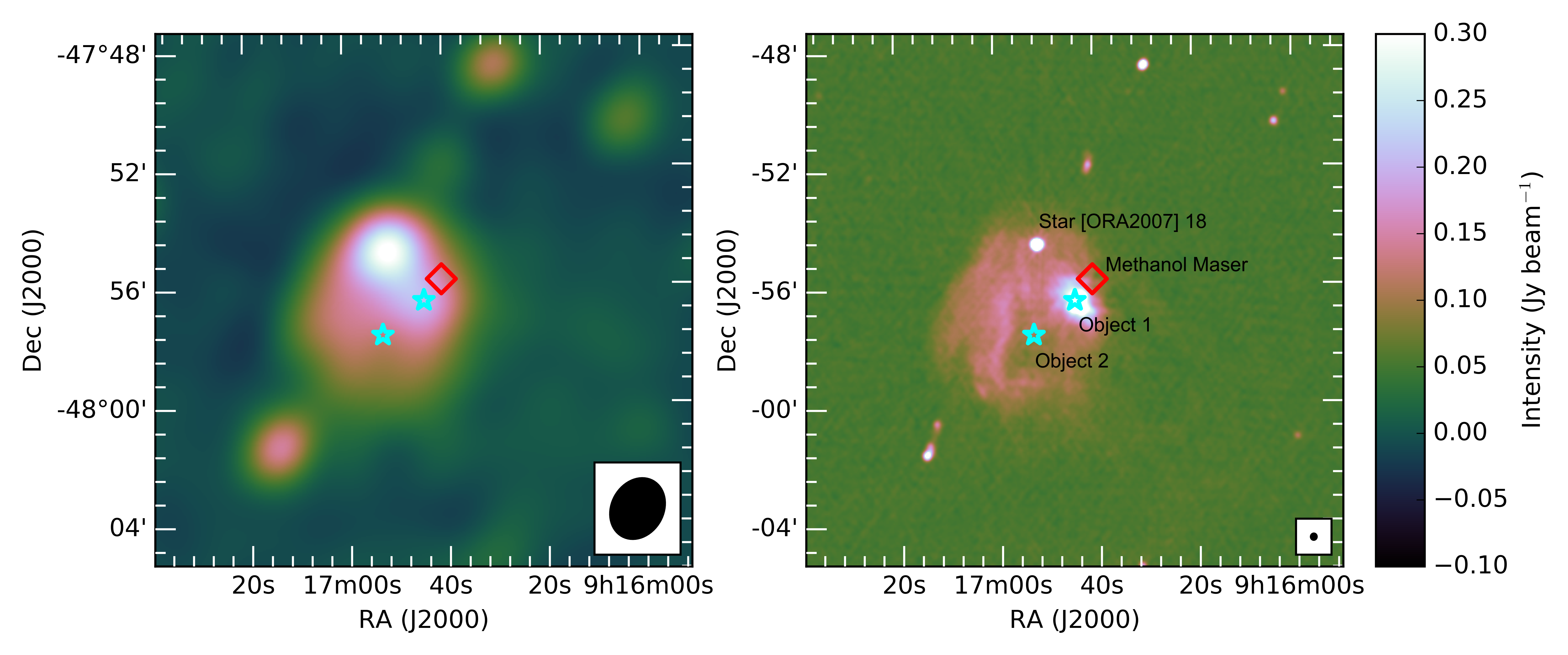}
\caption{RCW41 H{\sc ii} imaged in continuum with the MWA (left) at 114\,MHz and ASKAP (right) at 887.5\,MHz.  The ``*'' symbol marks unique source positions listed in the WISE HII catalogue.  The red diamond marks the location of the maser detections. The bottom right-hand corner shows the size and shape of the synthesised beam for each radio interferometer.}
\label{RCW41}
\end{figure*}

Within the infrared cloud IRAS09149-4743, the H{\sc ii} region RCW41 (also known as TYC 8170-333-1, and BRAN 246) is situated at RA 09:16:58, Dec --47:57:47 (J2000), embedded in the Vela Molecular Ridge. This cluster is shown to be part of Cloud A \citep{Murphy_1991} with two distinct regions; Object 2 to the south fuelled by a B0V star at a  distance of 1.2 $\pm$ 0.12\,kpc and Object 1 fuelled by a O9V star at 1.27 $\pm$ 0.13\,kpc \citep{Roman-Lopes_2009}. These two regions are marked by ``*'' on Figure \ref{RCW41}.  \cite{Santos_2012} studied the polarimetric properties of Object 1 and the sources within the cluster with deep infrared observations.  They determined the cluster age of 2.5--5\,MYr using a set of pre-main-sequence isochrones. In the analysis completed by \cite{Santos_2012}, they found several sources exhibiting infrared excess consistent with pre-main-sequence stellar objects and suggests there is a O9V star in Object 1, where the main cluster of stellar objects are found. They also suggest the extinction gradient in the south–north direction may suggest the O9V star, possibly the most massive in the cluster, has already swept out dust and gas.       

Object 1 shows a strong filament structure when viewed in H$\alpha$ and matches with the magnetic field directions reported by \cite{Santos_2012}. This region (G270.2614 +00.8358) also includes strong masers prominently detected, which all represent sign-post for high-mass star formation activity.  This has included water masers \citep{Breen_2011} at 22\,GHz, Class I and II methanol masers at 6.7, 36 and 44\,GHz (i.e \citealt{Green_2012,Voronkov_2014, Walsh_1997}) but no detection of hydrogen recombination lines were noted by \cite{Wenger_2021}, also suggesting the high-mass star is in the late stages of star formation. The LSR velocity for the 6.7\,GHz of 3.9km/s could be within 2.8km/s of the systemic (with reference there for \citealt{Green_2011}) and the kinematic distance was not determined on the basis of H{\sc i} self absorption, making it more ambiguous. 

\cite{Neichel_2015} studied both Objects in the region of RCW41 in detail using the Gemini-GeMS/GSAOI instrument for wide-field optic infrared. They find an age gradient along the cluster within and found 1/3 of the young stellar objects (YSOs) are in a range between 3 to 5\,Myr, while 2/3 of the YSO are $<$3/,Myr old. They also measured the total cluster mass of 78$\pm$18\,M$_{\odot}$ and that the initial mass function matches that of the Trapezium (around the Orion Nebula).  

Emission of H$\alpha$ and 5.8\,$\mu$m infrared suggests Object 1 is dominated by warm gas \citep{Neichel_2015}. The MWA and ASKAP data shows a concentrated region of emission toward the North that matches the WISE emission in Figure \ref{WISE_ASKAP} and matches the source structure commented by \cite{Santos_2012}. The high-resolution RACS image shows detailed structure in the emission which matches well with the source positions within the WISE Catalogue, marked with ``*'' on Figure \ref{RCW41}.  The MWA image is confused with emission from an (likely) unrelated star to the north of the region.

\subsection{GAL281.01--01.53}
This H{\sc ii} region, located at RA 09:59:08.5, Dec --56:52:44 (J2000) was first observed in the survey by \cite{Caswell_1987} and estimated to be at a distance of 1.9\,kpc.  The H$_{2}$CO velocity was measured at --7\,km\,s$^{-1}$ and the hydrogen RRLs at 4.8\,GHz at --5\,km\,s$^{-1}$ and they noted that there were no optical counterparts detected. \cite{Kuchar_1997} observed 760 H{\sc ii} regions with Parkes 64m and National Radio Astronomy Observatory Green Bank 100\,m telescopes and compared their flux densities at 4.85\,GHz with the those from \cite{Caswell_1987}.  At these frequencies the flux density was measured to be 8.2\,Jy.

\begin{figure}
\centering
\includegraphics[width=0.48\textwidth]{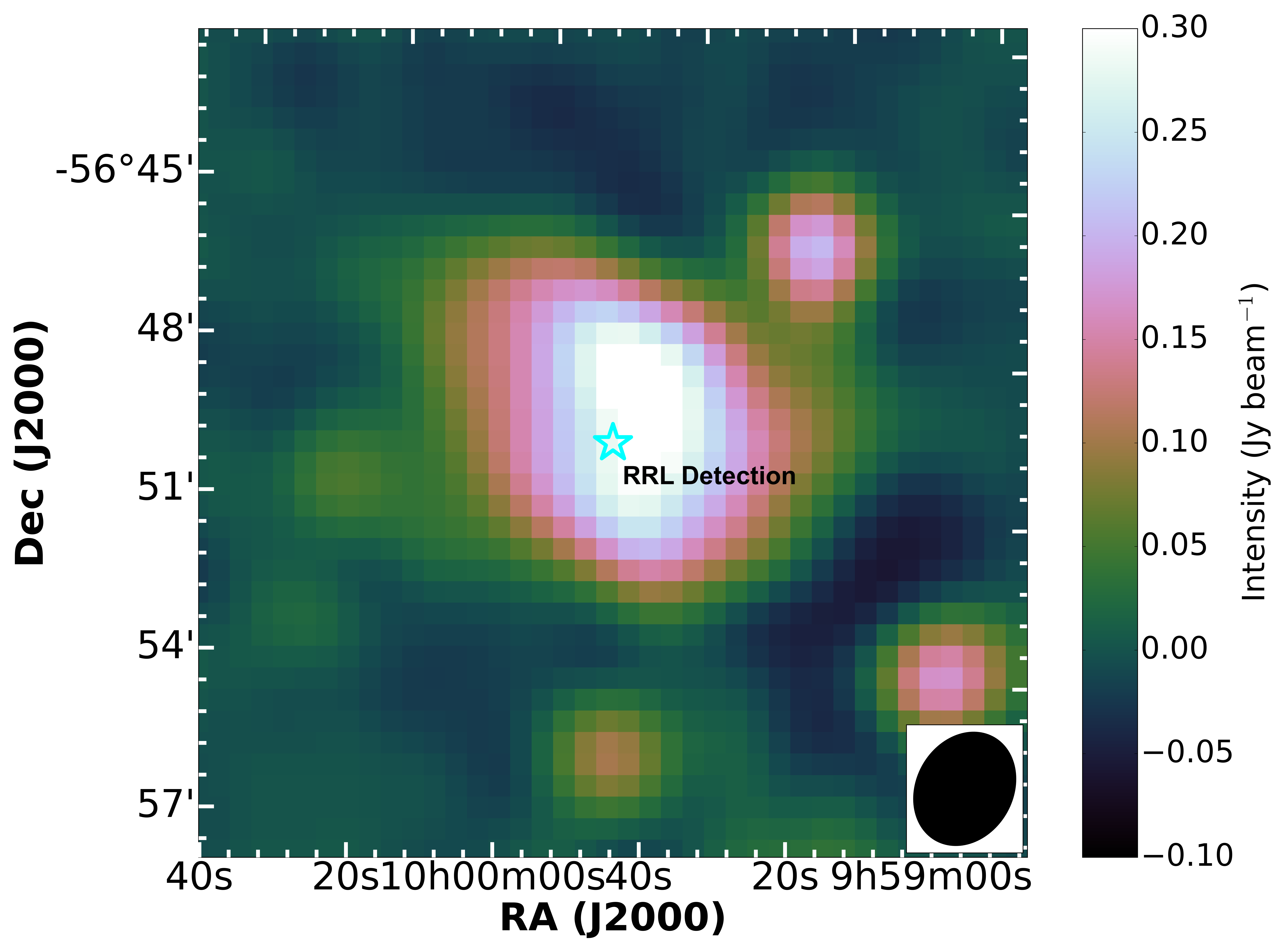}
\caption{GAL281.01--1.53 H{\sc ii} imaged in continuum with the MWA at 114\,MHz   The ``*'' symbol marks unique source positions listed in the WISE HII catalogue. The bottom right-hand corner shows the size and shape of the synthesised beam for each radio interferometer.} 
\label{GAL281}
\end{figure}

In Figure \ref{GAL281} the ``*'' marks the position of the RRL detection by \cite{Wenger_2021} and $^{13}$CO by \cite{Urquhart_2007}, which had velocities of --6.8 and --8.7\,km\,s${^-1}$ respectively. Although the source was included in the region of methanol maser surveys by \cite{Walsh_1997} and \cite{Van_1995}, no emission was detected. The MWA continuum image is only slightly resolved with the emission slightly to the left of the RRL source position.  The source was on the boarder of two RACS footprints and a high resolution image was not obtained. In the IRAS survey, emission was not detected at all frequencies so an upper limit on the luminosity was not determined.

\section{Discussion}
\cite{Yamaguchi_Gum_1999} observed the Gum Nebula in J=1--0 $^{12}$CO and found 82 molecular clouds and a number cometary globules thought to be nebula. Observations of CO trace the molecular gas associated with structures within giant molecular clouds and the study of the kinematics or velocity structures allows for an assessment of the population density of nebulae, which may contain H{\sc ii} regions. They determined that the velocity of these clouds differed from that of Galactic rotation and had a high velocity dispersion; suggesting the clouds are expanding under the dynamical motion caused by high-mass stars and supernova explosions, making it an interesting region to study with wide-field surveys.  It may also suggest that by combining these surveys, a significant number of new H{\sc ii} regions may be uncovered.

Radio astronomy surveys, in particular of those less than 1\,GHz, provide important detail on the non-thermal emission toward high-mass star forming regions.  Often, surveys done at high resolution are limited in field-of-view due to the large quantities of observational time and data processing required for wide-field analysis. However, with Square Kilometre Array (SKA) pathfinders, such as ASKAP and the MWA, we are learning the scope of science that can be accomplished with these new, high-resolution, wide-field telescopes.  

Due to a number of long-running radio astronomy programs and telescopes in the northern hemisphere, the region above Declination --20 is reasonably well characterised for known sources to help study star formation models. This survey is a pilot study to help build a database of southern hemisphere sources that could broaden our knowledge of star formation and important links in evolutionary phases which could be further explored as the next generation southern telescopes come online, like the Square Kilometre Array.

\begin{figure}
\centering
\includegraphics[width=0.48\textwidth]{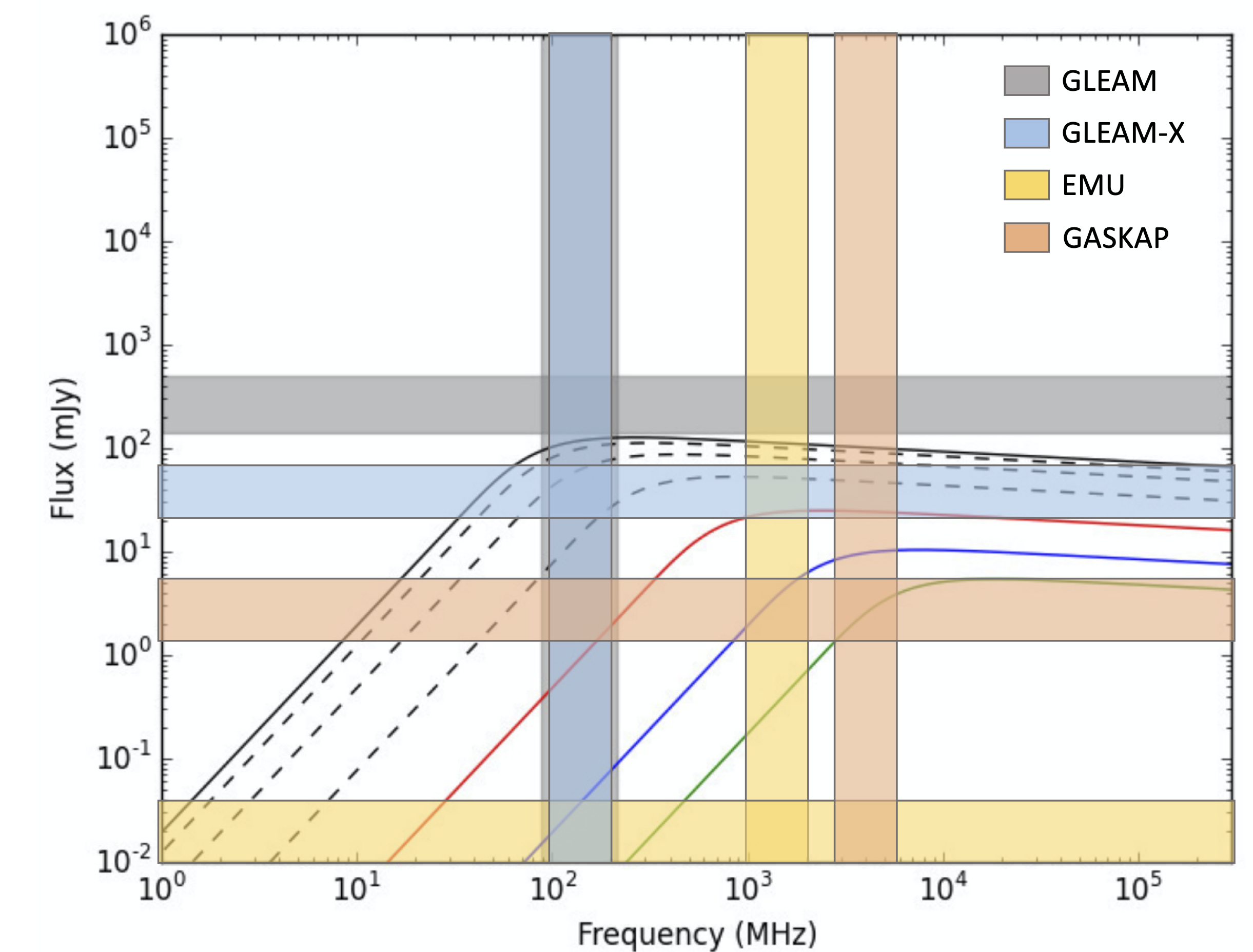}
\caption{Figure 5 from \citealt{Hindson_2016} showing model spectral energy distributions (SEDs) for different types of H{\sc ii} regions at a distance of 20\,kpc, with the expected detection ranges of upcoming surveys; The Galactic and Extra-galacitc All-sky MWA eXtended Survey (GLEAM-X; blue bars), Evolutionary Map of the Universe (EMU; Yellow bars) and Galactic ASKAP Survey (GASKAP; Orange bars). The green, blue, and red curves show model SEDs for hypercompact, ultracompact, and compact HII regions.  The black (solid and dashed) curves show model SEDs for classical HII regions of different sizes.  As shown, the next generation ASKAP surveys will be sensitive to the hypercompact regions (green line) out to 20\,kpc.}
\label{Other_Surveys}
\end{figure}

As shown in Figure \ref{Other_Surveys}, we are entering into a new phase in radio astronomy with an abundance of all-sky, high-resolution, low-frequency surveys that will allow us to study star formation regions to the further reaches of our Galaxy. The plot from Figure 5 in \cite{Hindson_2016}, shows the expected SEDs for compact to ultra compact H{\sc ii} regions for sources out to 20\,kpc based on equations 1–4 of \cite{1967ApJ...147..471M}. In their paper, they included the 5$\sigma$ limits for the GLEAM survey \citep{Hurley-Walker_2017}.  We have added the target sensitivity limits for upcoming surveys expected to be completed around 2021--2023. This includes the Evolutionary Map of the Universe (EMU;\citealt{Norris_EMU_2011,Norris_Emu_2021}), the GLEAM extended survey (GLEAM-X\footnote{https://www.mwatelescope.org/gleam-x}), and the Galactic ASKAP Survey (GASKAP;\citealt{Dickey_GASKAP}). The fully realised RACS survey (including mid and high-band ASKAP data;\citealt{McConnell_2020}) will also help in this analysis. The vertical bands represent the expected frequency range for these surveys and the horizontal bands represent the expected sensitivity (line width for the sensitivity are arbitrary). 

\section{Conclusion}
In this paper we provide a new multi-wavelength analysis and low-frequency information on RCW 38 and a selection of H{\sc ii} regions observable in the southern hemisphere. Additionally, we use the wide-field capabilities to simultaneously study the sources with ASKAP and MWA.  With new upcoming surveys like Evolutionary Map of the Universe (EMU;\citealt{Norris_EMU_2011}), the completion of the RACS survey along the ASKAP frequency band, and the upcoming extension to the Galactic and Extra-Galactic All-sky MWA Survey (GLEAM;\citealt{Hurley-Walker_2017}), many more surveys of this style will be possible, along with sensitivity to a broader range of H{\sc ii} regions.  

Many of these sources are interesting as they represent young H{\sc ii} regions that are densely packed with high-mass stars. Orion has long been a region of intense interest for testing models due to its close proximity and large number of high-mass stars.  However, to truly test the models, a larger population of sources need to be identified. This work provides a glimpse of what can be achieved with these larger surveys but also what is possible with the future Square Kilometre Array, in regards to unprecedented sensitivity and survey speeds.

\section*{Acknowledgements}

This publication makes use of data products from the Wide-field Infrared Survey Explorer, which is a joint project of the University of California, Los Angeles, and the Jet Propulsion Laboratory/California Institute of Technology, funded by the National Aeronautics and Space Administration.  This scientific work makes use of the Murchison Radio-astronomy Observatory, operated by CSIRO. We acknowledge the Wajarri Yamatji people as the traditional owners of the Observatory site. Support for the operation of the MWA is provided by the Australian Government (NCRIS), under a contract to Curtin University administered by Astronomy Australia Limited. The Australian SKA Pathfinder is part of the Australia Telescope National Facility which is managed by CSIRO. Operation of ASKAP is funded by the Australian Government with support from the National Collaborative Research Infrastructure Strategy. Establishment of ASKAP, the Murchison Radio-astronomy Observatory and the Pawsey Supercomputing Centre are initiatives of the Australian Government, with support from the Government of Western Australia and the Science and Industry Endowment Fund. We acknowledge the Pawsey Supercomputing Centre which is supported by the Western Australian and Australian Governments.  The Australia Telescope Compact Array is part of the Australia Telescope National Facility which is funded by the Australian Government for operation as a National Facility managed by CSIRO. We acknowledge the Gomeroi people as the traditional owners of the Observatory site.

For this work, we used the following software:
\begin{itemize}
    \item {\sc aoflagger} and {\sc cotter} -- \cite{OffriingaRFI}
    \item {\sc wsclean} -- \cite{offringa-wsclean-2014,offringa-wsclean-2017}
    \item {\sc Aegean} -- \cite{Hancock_2018_Aegean}
    \item {\sc topcat} -- \cite{Topcat}
    \item{{\sc NumPy} v1.11.3 \citep{NumPy}, {\sc AstroPy} v2.0.6 \citep{Astropy}, {\sc SciPy} v0.17.0 \citep{SciPy}, {\sc Matplotlib} v1.5.3 \citep{Matplotlib}}
    \item {\sc CARTA} -- \cite{angus_comrie_2020_3746095}
\end{itemize}

\section{Data Availability}
A description and citation to relevant survey papers, as available, are listed in Section 2 of this paper.

The MWA \& ATCA data underlying this article will be shared on reasonable request to the corresponding author. The raw visibilities for the MWA are available to registered users via the Murchison Widefield Array All-Sky Virtual Observatory at https://asvo.mwatelescope.org/. 

The WISE, AKARI, and SHASSA data underlying this article are available in $SkyView$ at https://skyview.gsfc.nasa.gov/current/cgi/titlepage.pl, and can be accessed with the relevant survey and wavelength name.

The WHAM survey data underlying this article are available in the University of Wisconsin archive at http://www.astro.wisc.edu/wham-site/wham-sky-survey/wham-ss/ via their web interface.

The Herschel Space Observatory data used in this article are available on the NASA/IPAC Infrared Science Archive at https://irsa.ipac.caltech.edu/ and is available on this web service.

The ASKAP RACS Survey data used in this article are available through the CSIRO ASKAP Science Data Archive at https://research.csiro.au/casda/ and is available to all registered users on the archive.

\bibliographystyle{mnras}
\bibliography{Vela_Survey} 

\appendix
\section{Calculations}
\subsection{Luminosity from IRAS Fluxes}
To estimate the upper limit on the luminosity we use the IRAS fluxes with equation 3 from \cite{Walsh_1997} which is modified from \cite{Casoli_1986}. The IRAS measurements can be considered as a strong upper limit on the bolometric luminosity as the large IRAS beam encapsulates all emission and there is no loss of flux due to
extended emission. 

\begin{equation}
    F_{tot} = (f_{12}\delta\nu_{25} + f_{25}\delta\nu_{12} + f_{60}\delta\nu_{60} + f_{100}\delta\nu_{100})/0.61
\end{equation}

where $f_{x}$ is the flux measured by Infrared astronomical satellite (IRAS) catalogs and atlases \citep{IRAS_PS} and $\delta\nu_{x}$ is the bandwidth for each band.  The value of 0.61 was derived by \cite{Walsh_1997} based on the analysis by \cite{Chini_1986} for compact H{\sc ii} regions. \cite{Tremblay_2015} found this upper limit to be comparable values determined by LABOCA and other infrared instruments. This equation produces different values than the Equation 1 from \cite{Helou_1985} used for estimating the total luminosity for extra-galactic sources.

An upper limit on the luminosity is then derived by:

\begin{equation}
    L_{\odot}=F_{tot} * 4 \pi r^{2}
\end{equation}

where r is the distance to the source in kpc. 

\subsection{Lyman Photon Flux}
To determine the Lyman Photon Flux values we are using the equations by \cite{Mezger_1967} but with the modifications and assumptions from equations 1 and 3 in \cite{Kurtz_1994}. 

\begin{equation}
    N_\mathrm{c} = 8.04\times10^{46} T_{\mathrm{e}}^{-0.85} U^3
\end{equation}

In this equation $T_{\mathrm{e}}$ is the electron temperature in Kelvins, which a value of 10$^{4}$ from \cite{Haynes_1978}.  The value $U$ is derived as:

\begin{equation}
    U = 4.553 [\frac{1}{\alpha(\nu,T_{\mathrm{e}})} (\frac{\nu}{GHz})^{0.1} (\frac{T_{\mathrm{e}}}{K})^{0.35} (\frac{S_\nu}{Jy}) (\frac{D}{kpc})^2]^{\frac{1}{3}}
\end{equation}

For these values in Table 1, we used the flux density and frequencies from the ASKAP data, as the surface brightness sensitivity is better than the MWA with the short baselines removed.

\subsection{Data Availability}
A description and citation to relevant survey papers, as available, are listed in Section 2 of this paper.

The MWA \& ATCA data underlying this article will be shared on reasonable request to the corresponding author. The raw visibilities for the MWA are available to registered users via the Murchison Widefield Array All-Sky Virtual Observatory at https://asvo.mwatelescope.org/. 

The WISE, AKARI, and SHASSA data underlying this article are available in $SkyView$ at https://skyview.gsfc.nasa.gov/current/cgi/titlepage.pl, and can be accessed with the relevant survey and wavelength name.

The WHAM survey data underlying this article are available in the University of Wisconsin archive at http://www.astro.wisc.edu/wham-site/wham-sky-survey/wham-ss/ via their web interface.

The Herschel Space Observatory data used in this article are available on the NASA/IPAC Infrared Science Archive at https://irsa.ipac.caltech.edu/ and is available on this web service.

The ASKAP RACS Survey data used in this article are available through the CSIRO ASKAP Science Data Archive at https://research.csiro.au/casda/ and is available to all registered users on the archive.

\bsp	
\label{lastpage}
\end{document}